\documentclass[11pt, a4paper]{googlecloud}

\pdftrailerid{}
\usepackage[authoryear, sort&compress, round]{natbib}

\usepackage{url}
\usepackage{graphicx}
\usepackage{booktabs}
\usepackage[most]{tcolorbox}
\usepackage{multirow}
\usepackage{algorithm}
\usepackage[noend]{algorithmic}

\usepackage{colortbl}
\usepackage{array}
\usepackage{caption}
\usepackage{amssymb}
\usepackage{enumitem}
\usepackage{etoolbox}
\apptocmd{\ttfamily}{\hyphenchar\font=`\-}{}{}
\definecolor{GoogleBlue}{HTML}{4285F4}
\definecolor{GoogleRed}{HTML}{EA4335}
\definecolor{GoogleYellow}{HTML}{FBBC05}
\definecolor{GoogleGreen}{HTML}{34A853}

\usepackage{bm}
\usepackage{xspace}
\usepackage{tikz}
\usepackage{wrapfig}
\usepackage{adjustbox}
\usepackage{subcaption}
\usepackage{tabularx}
\usepackage{soul}
\usepackage{arydshln}

\title{\sys: Learned Context Curation for LLM Catalog Search}

\correspondingauthor{\footnotesize{Corresponding author(s): chuxuan3@illinois.edu, \{hejiec, normanyhuang, skbharti, wangchiew, soarik\}@google.com} \\ \textup{\footnotesize{* This work was done while Chuxuan was a Student Researcher at Google.}}}

\author[1, 2*]{Chuxuan Hu}
\author[1]{Hejie Cui}
\author[1]{Norman Huang}
\author[1]{Shubham Kumar Bharti}
\author[1]{Wang-Chiew Tan}

\author[1]{\mbox{Sercan \"{O}. Ar{\i}k}}

\affil[1]{Google}
\affil[2]{University of Illinois Urbana-Champaign}

\newcommand{\sys}{\textsc{RPTune}\xspace}
\definecolor{trainable_module}{HTML}{D6B656}
\definecolor{StepBadge}{HTML}{6A8791}
\newcommand{\stepnum}[1]{%
\tikz[baseline=(char.base)]{
\node[shape=circle, fill=StepBadge, text=white,
      font=\small, inner sep=1pt] (char) {#1};}}
\newcommand{\minihead}[1]{\vspace{.1em}\noindent\textbf{#1.}}

\definecolor{prelimgray}{gray}{0.55}

\definecolor{AllBlue}{RGB}{52,73,94}
\definecolor{AllHighlight}{RGB}{235,242,250}

\definecolor{SysRowBlue}{RGB}{235,242,250}
\definecolor{SysGainBlue}{RGB}{44,95,140}
\definecolor{SysGainGreen}{RGB}{21,104,66}
\definecolor{AvgEMTint}{HTML}{F4F0E8}
\definecolor{AvgEMTintSys}{RGB}{214,229,245}
\definecolor{DefaultHL}{HTML}{F2E3CF}
\newcommand{\dflt}[1]{\smash{\setlength{\fboxsep}{1.2pt}%
  \colorbox{DefaultHL}{#1}}}
\newcommand{\sharedbackbone}{%
  \begingroup
  \setlength{\fboxsep}{0.7pt}%
  \colorbox{black!8}{\textnormal{gemini-3.7-flash}}%
  \endgroup
}

\newcommand{\smbP}[1]{\,\textcolor{SysGainBlue}{\footnotesize($\bm{\uparrow}$#1)}}
\newcommand{\smbPd}[1]{\,\textcolor{SysGainGreen}{\footnotesize($\bm{\downarrow}$\textbf{#1})}}
\newcommand{\smbPB}[1]{\,\textcolor{SysGainBlue}{\footnotesize($\bm{\uparrow}$\textbf{#1})}}

\definecolor{PLBlue}{HTML}{A3CFDF}
\definecolor{PLav}{HTML}{D0D1E6}
\definecolor{PTaupe}{HTML}{CDBAB7}
\definecolor{QContext}{HTML}{FFF5CC}

\colorlet{QRequirement}{PLBlue!55}
\colorlet{QPreference}{PLav!65}

\newcommand{\qreq}[1]{{\sethlcolor{QRequirement}\hl{#1}}}
\newcommand{\qpref}[1]{{\sethlcolor{QPreference}\hl{#1}}}
\newcommand{\qcontext}[1]{{\sethlcolor{QContext}\hl{#1}}}

\newcounter{exquery}
\renewcommand{\theexquery}{Q\arabic{exquery}}

\fancypagestyle{firststyle}{%
  \fancyhead[L]{\includegraphics[width=60pt]{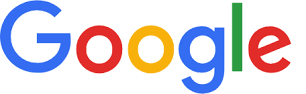}}%
  \fancyhead[C,R]{}%
  \fancyfoot[L]{\footerfont\itshape\the\correspondingauthor \\
    \footerfont\textcopyright\, \the\year{} Google. All rights reserved}%
  \fancyfoot[C,R]{}%
}

\begin{abstract}
For small merchant businesses (SMBs) whose catalogs fit within
a long-context LLM, full-catalog prompting offers a compelling alternative
to multi-stage retrieval designed primarily for large marketplaces with millions of items. However, fitting the catalog does not
ensure that the model can use it effectively: LLMs do not utilize
long contexts uniformly. We therefore study \emph{in-context
catalog search} through two complementary questions:
(1) how to curate and present catalogs to the LLM, and
(2) how to adapt the LLM for product selection on curated contexts.

We propose \sys, an end-to-end framework that couples learned catalog curation with LLM post-training using automatically generated, catalog-grounded supervision.
An encoder--reorganizer
ranks, prunes, and positions products guided by downstream
LLM feedback, while the resulting curated catalogs in turn improve the effectiveness of LLM post-training with a context-relative reward.
We evaluate \sys on 7 real merchants spanning distinct retail verticals, using 100 complex conversational queries per merchant. \sys consistently improves search accuracy across both proprietary and open-weight LLMs, with context curation yielding gains of up to 31.4 percentage points and post-training adding a further 10.3 points on average.
\end{abstract}

\begin{document}

\maketitle

\section{Introduction}
\label{sec:intro}

\begin{wrapfigure}{r}{0.495\textwidth}
\centering
\vspace{-3mm}
\includegraphics[width=0.495\textwidth]{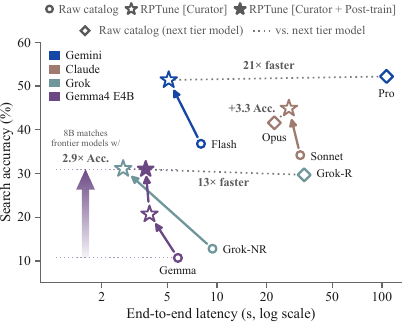}
\newcommand{\glyphlink}{\raisebox{0.22ex}{\includegraphics[width=1.15em]{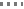}}}
\newcommand{\glyphbetter}{\raisebox{-0.55ex}{\includegraphics[height=2.5ex]{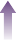}}}
\vspace{-3mm}
\caption{Search accuracy vs.\ end-to-end latency, averaged
over 7 merchants; complete results and illustrations can be found in Table~\ref{tab:main_results}. \sys{} context curation delivers accuracy
gains comparable to a model upgrade with
up to 21$\times$ speedup (dashed cross-model comparisons in \mbox{\glyphlink}), while \sys post-training further improves \texttt{gemma-4-E4B-it} accuracy to comparable levels of frontier models \mbox{\glyphbetter}.}
\label{fig:teaser}
\end{wrapfigure}

Small merchant businesses (SMBs) are widespread online: Shopify alone hosts over 3 million active storefronts as of September 2026~\citep{storeleads_shopify_2026}. 
However, existing product search methods are developed primarily for large marketplaces operating over millions of items with dedicated machine learning teams and abundant behavioral signals. Because these assumptions do not hold for SMBs, traditional multi-stage retrieval pipelines are unnecessarily complex and costly for their catalog scales.

Long-context LLMs create a new opportunity in this regime. Among the 56 real-world Shopify storefronts we analyze, 92.9\% of catalogs fit within 1M tokens, 
with a median catalog size of only 51K tokens. 
This enables a compelling alternative to conventional retrieval-based search: instead of first retrieving a small candidate set, an LLM can directly reason over the full catalog. Specifically, we find that full-catalog LLM search improves accuracy over prior retrieval-based approaches by up to 20 percentage points.

However, simply placing the full catalog into an LLM context leaves substantial room for improvement. First, the catalog itself can be curated and organized to better support LLM reasoning. Prior work has shown that LLMs are sensitive to both the ordering and length of their input context, with relevant information placed in the middle of a long context often used less effectively than information near its beginning or end~\citep{liu2023lostmiddlelanguagemodels}. Second, the LLM itself can be adapted to better interpret these curated contexts for product search.

\begin{figure*}[t]
    \centering
    \includegraphics[width=\textwidth]{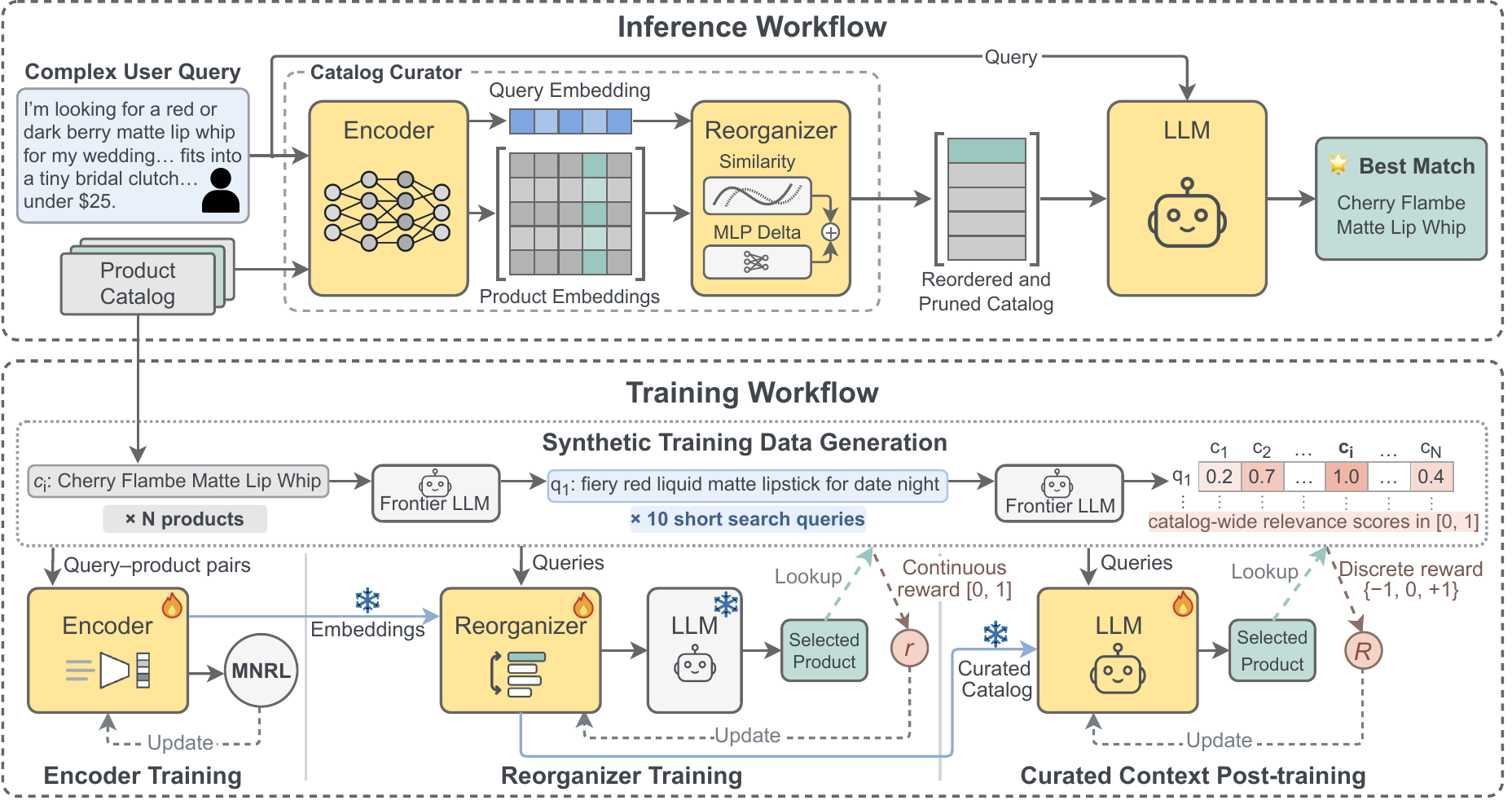}
\newcommand{\glyphtrainable}{\raisebox{-0.25ex}{\includegraphics[height=2.3ex]{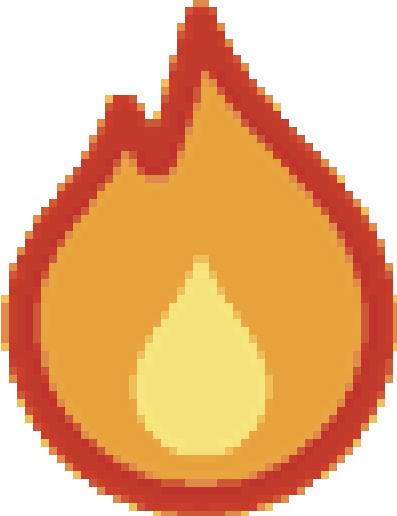}}}
\newcommand{\glyphfrozen}{\raisebox{-0.25ex}{\includegraphics[height=2.1ex, width=2.1ex]{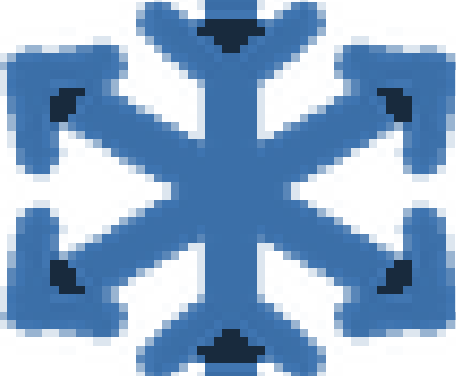}}}
\caption{\textbf{\sys architecture.} \underline{\textbf{Top:}} At inference, the curator (encoder + reorganizer) orders and prunes the catalog to construct a compact context for downstream LLM product selection. \underline{\textbf{Bottom:}} Synthetic queries and relevance scores support
sequential encoder training, reorganizer training with
frozen-LLM feedback, and LLM post-training on curated contexts.
\mbox{Flames\,\glyphtrainable} /\mbox{snowflakes\,\glyphfrozen} indicate trainable/frozen modules at different stages.}
    \label{fig:architecture}
\end{figure*}

We propose \sys, a framework for in-context catalog search
that combines learned context curation with LLM post-training in Section~\ref{sec:method}.
As we show in Figure~\ref{fig:architecture},
a curator with encoder--reorganizer structure prunes and orders products into compact,
position-aware contexts, with the reorganizer trained using
downstream LLM feedback. We then post-train the LLM on these
curated contexts to select the best available product.
All training uses synthetic, catalog-grounded supervision,
requiring no merchant-provided relevance labels.

We evaluate \sys on 7 real merchants spanning distinct retail verticals with 100 complex conversational queries per merchant. Across 8 evaluated frontier LLMs, \sys's context curation consistently improves search accuracy for all merchants, with gains of 14.0 percentage points on average and up to 31.4 points. These gains extend to open-weight LLMs: the full \sys pipeline improves \texttt{gemma-4-E4B-it} search accuracy by 20.3 points on average, with post-training alone contributing an additional 10.3-point average gain. We report our detailed findings in Section \ref{sec:experiment_results}.

\section{Problem: In-context Catalog Search}
\label{sec:usecase}

\minihead{Use case}
Consider the following query for
\textit{Beauty Bakerie} (Table \ref{tab:data}),
a cosmetics merchant:

\begin{quote}
\small
\refstepcounter{exquery}\label{q:wedding}%
\textbf{\theexquery.} ``I'm looking for a
\qreq{completely smudge-proof, long-wearing red or dark berry
matte lip whip} for \qcontext{my wedding day} that will survive
\qcontext{a 4-course meal and lots of champagne}
without budging! I'd love it to be
\qreq{100\% vegan and cruelty-free}.
Also, since I'm carrying a \qcontext{tiny bridal clutch},
the product needs to be super lightweight,
\qpref{ideally under 10 grams}, and
\qpref{I'm hoping to keep it under \$25}.''
\end{quote}

\ref{q:wedding} combines \qreq{explicit requirements} on color, finish,
wear performance, and vegan and cruelty-free attributes with
\qpref{flexible preferences} on weight and price, signaled
by ``ideally'' and ``hoping.''
It also includes \qcontext{contextual cues}: the bridal clutch
implies a preference for compact packaging, while the wedding
and meal scenario conveys expectations about durability.
Answering therefore requires interpreting conversational
language, jointly evaluating interacting constraints, and
considering trade-offs when no variant satisfies them all. 

Addressing these requirements demands a holistic view of the catalog, and is directly feasible for SMBs: Beauty Bakerie's entire inventory contains around 100 product variants and fewer than 40K tokens, fitting completely within a modern LLM's context window. Thus, feeding the full catalog directly into the prompt allows LLM to jointly reason over all items at once and directly select the best-matching product.

\minihead{Problem setting}
We study \emph{in-context catalog search}.
Given a single merchant's catalog
$\mathcal{C}=\{c_1,\ldots,c_n\}$ of product variants and a
single-turn natural-language query $q$, the task is to select
the best-matching variant $\hat{c}\in\mathcal{C}$ based on the
available catalog metadata.
Figure~\ref{fig:architecture} illustrates an example workflow
for \ref{q:wedding}. 
In our setting, \underline{the catalog is the only merchant-provided data source required,
including for training} (no merchant-provided relevance labels
or user behaviour logs are needed).
We focus on catalogs that fit within the LLM's context window
alongside the query and instructions.

\section{Method: \sys}
\label{sec:method}

We propose \sys (\textbf{R}ank, \textbf{P}rune, and Fine\textbf{tune}), a framework for in-context product search
that combines a \emph{catalog curator} with a downstream LLM
(Figure~\ref{fig:architecture}).
We illustrate the inference and training workflows in Sections~\ref{subsec:workflow} and~\ref{subsec:training}, respectively.

\subsection{\sys Inference Workflow}
\label{subsec:workflow}

Given a user query $q$ and a product catalog
$\mathcal{C}=\{c_1,\dots,c_n\}$, \sys predicts the
best-matching product
$\hat{p} = \text{\sys}(q,\mathcal{C}) \in \mathcal{C}.
$
Its catalog curator comprises an \emph{encoder}, which
maps the query and products into dense representations,
and a \emph{reorganizer}, which uses these representations
to select and arrange products into a compact,
query-dependent context $\widetilde{\mathcal{C}}_q$
for the downstream LLM.

The encoder maps the query and each product into
$L_2$-normalized dense representations:
$$
\mathbf{e}_q = \text{Encoder}(q), \quad \mathbf{e}_{c_i} = \text{Encoder}(c_i) \quad \forall c_i \in \mathcal{C}.
$$

The reorganizer constructs a
compact, query-dependent catalog context based on these representations from the encoder:
\[
\widetilde{\mathcal{C}}_q =
\operatorname{Reorganizer}
\left(\mathcal{C}, \mathbf{e}_q,
(\mathbf{e}_{c_i})_{i=1}^{n}; \rho\right),
\]
where $\rho\in[0,1)$ is the pruning rate, specifying the
fraction of catalog products to discard.
The output $\widetilde{\mathcal{C}}_q$ is an ordered subset
containing $k=\lceil(1-\rho)n\rceil$ products.
The reorganizer determines both which products to retain
and where to place them in the context.

Internally, it assigns each product a context-priority score $s_i$
that combines embedding-based semantic matching with a
learned task-specific adjustment:
\[
s_i =
\underbrace{\tau \cdot
(\mathbf{e}_q^\top \mathbf{e}_{c_i})}_{\text{Base similarity}}
+
\underbrace{\operatorname{MLP}
([\mathbf{e}_q;\mathbf{e}_{c_i}])}_{\text{Learned adjustment}},
\]
where $[\cdot;\cdot]$ denotes vector concatenation,
$\operatorname{MLP}$ is a multilayer perceptron with a scalar
output, and $\tau$ controls the contribution of the base
similarity (we set $\tau=10$).
Let $\pi$ be a permutation satisfying
$s_{\pi(1)} \ge \dots \ge s_{\pi(n)}$.
We retain the top $k$ products and arrange them in
ascending order of priority, placing higher-scoring
products closer to the final generation prompt:
$
\widetilde{\mathcal{C}}_q =
(c_{\pi(k)}, c_{\pi(k-1)}, \dots, c_{\pi(1)}).
$
The downstream LLM then predicts the best-matching product
from the curated context:
$
\hat{p} = \operatorname{LLM}
\left(q,\widetilde{\mathcal{C}}_q\right).
$

\subsection{\sys Training Workflow}
\label{subsec:training}
\sys training proceeds sequentially using synthetic, catalog-grounded data; at each stage, only the target module is updated while the others remain frozen.

\minihead{Training Data Generation}
We generate synthetic query--product pairs $\mathcal{D}_{\text{pair}}$ and catalog-wide relevance scores $\mathcal{D}_{\text{rel}}$ in two steps. We record the details in Appendix \ref{app:training_data}.

\textit{Query--Product Pair Generation.} For each product variant $c_i$ in the merchant's catalog, we prompt a frontier LLM (\texttt{gemini-3.1-pro}) to generate 10 diverse, 1-2 sentence search queries targeting $c_i$. This yields $\mathcal{D}_{\text{pair}} = \{(q, c_{q})\}$, a set of $10n$ realistic queries paired with their target variants.

\textit{Catalog-Wide Relevance Scoring.} To capture partial matches and near-substitutes, we prompt the LLM to assign a graded relevance score $y(q, c) \in [0.0, 1.0]$ to every catalog variant $c$ for each query $q$. To maintain global consistency, scoring runs in batches of 10 variants per call alongside a running dictionary of previously assigned scores. This yields a dense relevance distribution $\mathcal{D}_{\text{rel}} = \{(q, c, y(q,c))\}$ over the entire catalog for every query.

\minihead{Encoder Training}
The encoder must capture the distinguishing features of each product and
place queries near their matching products in a shared semantic space. To achieve this, we train the encoder using the Multiple Negatives Ranking Loss (MNRL) \citep{henderson2017efficientnaturallanguageresponse}. 

Given a batch of $B$ training pairs $\{(q_i, c_i)\}_{i=1}^B \sim
\mathcal{D}_{\text{pair}}$, the encoder generates their L2-normalized dense embeddings $\mathbf{e}_{q_i} = \text{Encoder}(q_i)$ and $\mathbf{e}_{c_i} = \text{Encoder}(c_i)$. The objective increases the similarity of each positive pair
$(q_i,c_i)$ relative to the in-batch negatives
$(q_i,c_j)$ for $j\neq i$:
$$
\mathcal{L}_{\text{Encoder}} = - \frac{1}{B} \sum_{i=1}^B \log \frac{\exp\left(\mathbf{e}_{q_i}^\top \mathbf{e}_{c_i}\right)}{\sum_{j=1}^B \exp\left(\mathbf{e}_{q_i}^\top \mathbf{e}_{c_j}\right)}
$$

\begin{algorithm}[t]
\footnotesize
\caption{\sys Reorganizer Training via LLM-in-the-Loop RL}
\label{alg:reorganizer_rl}
\begin{algorithmic}[1]
\REQUIRE Catalog $\mathcal{C}$; relevance data
$\mathcal{D}_{\text{rel}}=\{(q,c,y(q,c))\}$,
where $y(q,c)\in[0,1]$ is the pre-computed relevance
of product $c$ to query $q$; frozen encoder and downstream LLM
\STATE Initialize $\theta \gets$ zero-initialized output layer.
\FOR{step $= 1, \ldots, T$}
    \STATE Sample a batch $\mathcal{B}$ of $B$ queries from $\mathcal{D}_{\text{rel}}$
    \FOR{each query $q \in \mathcal{B}$}
\STATE Compute context-priority scores $\{s_i\}_{i=1}^{|\mathcal{C}|}$ and construct curated
$\widetilde{\mathcal{C}}_q$ following
Section~\ref{subsec:workflow}.
        \STATE Sample $G$ selections
        $a_g\sim\operatorname{LLM}
        (q,\widetilde{\mathcal{C}}_q)$.
        \STATE For each $g$, retrieve
        $r_g\gets y(q,a_g)$ from $\mathcal{D}_{\text{rel}}$
        if $a_g\in\mathcal{C}$; otherwise, set $r_g\gets -1$.
        \STATE Compute $A_g \gets (r_g-\mu_q)/(\sigma_q+\epsilon)$ from the group mean $\mu_q$ and std\ $\sigma_q$.
    \ENDFOR
 \STATE Update only $\theta$ using
    $\nabla_\theta\mathcal{L}_{\text{Reorganizer}}(\theta)$
    from Eq.~\ref{eq:reorganizer_loss}.
\ENDFOR
\ENSURE $\theta$
\end{algorithmic}
\end{algorithm}

\minihead{Reorganizer Training via LLM-in-the-Loop Reinforcement Learning}
The encoder places queries near semantically matching products, but its
pairwise training objective does not account for how the downstream LLM
actually behaves when presented with a curated catalog. We therefore train the reorganizer using
LLM feedback, as summarized in Algorithm~\ref{alg:reorganizer_rl}.
Only the reorganizer's MLP parameters $\theta$ are updated;
the encoder and downstream LLM remain frozen. The initial zero
MLP adjustment preserves the encoder-based ordering.

For each query $q$, the frozen LLM (\texttt{gemma-4-E4B-it})
produces $G$ single-product selections using nonzero-temperature
decoding from the same deterministic context
$\widetilde{\mathcal{C}}_q$.
Rewards are retrieved from $\mathcal{D}_{\text{rel}}$, with
$-1$ assigned to outputs outside $\mathcal{C}$.\footnote{
Note that
$a_g\in\mathcal{C}\setminus\widetilde{\mathcal{C}}_q$
still
receives the relevance rewards; positive advantages encourage
higher context priority, allowing the reorganizer to correct
potentially mistaken omissions.
} 
We standardize rewards within each query group to obtain
advantages $A_g$, following the group-relative normalization
of GRPO~\citep{shao2024deepseekmath}.
Groups with identical rewards yield zero advantages, so we design the optimization to operate over minibatches of $B$ queries, $\mathcal{B}=\{q_b\}_{b=1}^{B}$, where each update aggregates multiple query groups and includes informative groups with learning signals.

We use the LLM selections, weighted by their advantages, as the training signal for updating the reorganizer scores. Specifically, we define a catalog-wide distribution $w_\theta$ induced by these scores and minimize the surrogate objective $\mathcal{L}_{\text{Reorganizer}}$:
\begin{equation}
\label{eq:reorganizer_loss}
\begin{aligned}
w_\theta(c_i\mid q)
= \frac{\exp(s_i/\beta)}{\sum_{c_j\in\mathcal{C}}\exp(s_j/\beta)},
\qquad
\mathcal{L}_{\text{Reorganizer}}(\theta)
= -\frac{1}{B}\sum_{q\in\mathcal{B}}\frac{1}{G}
\sum_{g\in\mathcal{V}_q} A_g\log w_\theta(a_g\mid q).
\end{aligned}
\end{equation}

where $\beta$ is the temperature and
$\mathcal{V}_q=\bigl\{\,g\in\{1,\dots,G\} : a_g\in\mathcal{C}\,\bigr\}$ denotes the valid rollouts. Invalid outputs contribute to the
group statistics but are excluded from the loss. 

Note that $a_g$ is sampled from the frozen LLM rather than from $w_\theta$; $w_\theta$ provides a differentiable link from the LLM selections to the reorganizer scores. A positive advantage increases the score of the selected product relative to the rest of the catalog, while a negative advantage decreases it. Because these scores determine product inclusion and positioning in $\widetilde{\mathcal{C}}_q$, the update directly adjusts subsequent context construction. 
Every optimizer update contains informative groups with non-zero advantages, with detailed training statistics and curves reported in Appendix~\ref{app:curves}.

\minihead{LLM Post-training on Curated Catalogs}
After training the catalog curator, we freeze both the encoder
and reorganizer and post-train only the downstream LLM.
For each query $q$ in $\mathcal{D}_{\text{rel}}$, the frozen
curator constructs a query-dependent curated catalog
$\widetilde{\mathcal{C}}_q$.
We then use GRPO~\citep{shao2024deepseekmath} to sample
$G$ responses from the LLM and optimize its product selection
over $\widetilde{\mathcal{C}}_q$.

For each sampled response, let $c_{\mathrm{pred}}$ denote the
predicted product. 
Using the pre-computed relevance scores $y(q,c)$ from $\mathcal{D}_{\text{rel}}$, we define a context-relative reward: $R = 1$ if $c_{\mathrm{pred}} = \arg\max_{c \in \widetilde{\mathcal{C}}_q} y(q,c)$, $R = -1$ if $c_{\mathrm{pred}} \notin \widetilde{\mathcal{C}}_q$, and $R = 0$ otherwise.
The $-1$ case penalizes hallucination where the model hallucinates a product absent from
the curated catalog. The $+1$ case trains the LLM to select the highest-relevance
product available in the curated catalog, rather than requiring
the globally best product in the original catalog to be present.
This is important when catalog curation omits the globally best
variant: the LLM can still receive a positive reward for selecting
the best available alternative, preserving a useful learning
signal for GRPO.

\section{Experiments}
\label{sec:experiment_results}

We design our experiments around four questions:
\textbf{(RQ1)} Is full-catalog prompting a feasible and effective alternative to retrieval-based search for SMBs?
\textbf{(RQ2)} Does \sys's context curation improve search accuracy and efficiency across different LLMs?
\textbf{(RQ3)} Does post-training on curated contexts provide further accuracy gains beyond inference-time curation?
\textbf{(RQ4)} How well does \sys generalize to newly added products and unseen merchants?
\vspace{-3mm}
\begin{wraptable}{r}{0.5\textwidth}
\centering
\scriptsize
\renewcommand{\arraystretch}{0.95}
\setlength{\tabcolsep}{3pt}
\vspace{15mm}
\caption{Statistics of the 7 evaluated SMBs.}
\vspace{-2mm}
\label{tab:data}
{\tiny
\begin{tabular}{llrrr}
\toprule
\textbf{Merchant} & \textbf{Retail vertical} & \textbf{\# Prod.} & \textbf{\# Var.} & \textbf{\# Tok.} \\
\midrule
Beauty Bakerie (BB)          & Cosmetics            & 96  & 115 & 37{,}506 \\
Oats Overnight (OON)         & Breakfast food       & 142 & 303 & 39{,}845 \\
The Honest Kitchen (HK)      & Dog \& cat food      & 131 & 214 & 41{,}954 \\
Wilsun Custom Products (WCP) & Custom gears         & 110 & 410 & 81{,}355 \\
Wairua Beauty (WB)           & Skin \& hair care    & 277 & 409 & 84{,}870 \\
Focus \& Frame Eyewear (FF)  & Eyewear              & 186 & 739 & 110{,}724 \\
Package Free (PF)            & Zero-waste goods     & 116 & 278 & 126{,}958 \\
\bottomrule
\end{tabular}}
\vspace{-11mm}
\end{wraptable}

\subsection{Experiment Setup}
\label{subsec:setups}
\minihead{Evaluation Data}
We evaluate \sys and baselines on 7 real-world SMBs spanning distinct retail verticals (Table \ref{tab:data}). To avoid relying on sensitive user logs, we synthesize 100 complex conversational queries per merchant using a three-stage pipeline following prior work on LLM-generated evaluation data with human
quality checks \citep{yao2024taubenchbenchmarktoolagentuserinteraction, chen2023benchmarkinglargelanguagemodels, qin2023toolllmfacilitatinglargelanguage}. First, a multimodal LLM (\texttt{gemini-3.1-pro}) acts as a synthetic customer, generating multi-constraint queries based on high-level merchant context (e.g., ``About Us'' screenshots) while strictly withholding individual product records to prevent trivial keyword matching. 
Second, we use selection consensus across randomized catalog
orderings as a difficulty-control heuristic, retaining queries
with moderate consensus (31.9\% of candidates) to avoid trivially
easy or overly ambiguous cases.
Finally, human experts verify pilot queries and audit the final dataset to establish ground-truth labels  (resulting in a 9.6\% rejection rate).
Details in Appendix~\ref{app:data}.

\minihead{Models and Baselines}
We evaluate \sys across 9 LLM backbones from multiple model families.
For each frozen backbone, we compare raw full-catalog prompting (randomly shuffled) against \sys-curated contexts retaining the top 25\% of variants. To isolate curation quality, we compare the \sys curator against dense retrieval (EmbeddingGemma 300M, upon which the \sys encoder is trained, and Gemini Embedding 2) and listwise ranking (Jina Reranker 3.5) at the exact same 25\% retention budget. We post-train \texttt{gemma-4-E4B-it} (\texttt{gemma}) to evaluate the full \sys pipeline. Furthermore, using \sharedbackbone{} as a shared backbone, we compare against traditional search pipelines: standard RAG (top-10), RAG-Fusion~\citep{Rackauckas_2024,medrano2026scalingretrievalaugmentedgeneration}, TourRank~\citep{chen2025tourrankutilizinglargelanguage}, and LongLLMLingua~\citep{jiang2024longllmlinguaacceleratingenhancingllms}. Finally, Section~\ref{subsec:ablation_analysis} compares \sys's GRPO post-training objective against DPO~\citep{rafailov2024directpreferenceoptimizationlanguage} and IRPO~\citep{wu2025incontextrankingpreferenceoptimization}. We provide the full implementation details 
in Appendix \ref{app:setups}.

\minihead{Metrics}
For each SMB, we run each method 10 times on the same query
set and report averages over queries and runs.
We use exact-match accuracy (EM) as our primary effectiveness
metric, measuring the fraction of predictions that exactly match
the ground-truth product variant.
We additionally report feature reward (FR), the weighted percentage
of matching attributes between the selected and ground-truth variants,
to capture partial feature-level agreement
(formal definition in Appendix~\ref{app:metrics});
and end-to-end latency, measured from query submission to final
product selection.
We report EM and FR as percentages and latency in seconds.

\subsection{Main Results}
\label{subsec:main_res}

Table~\ref{tab:main_results} summarizes the main results for RQ1--RQ3 with full underlying results in
Appendix~\ref{app:full_results}.

\minihead{Full-catalog prompting outperforms retrieval-based baselines (RQ1)}
Using \sharedbackbone{} as the common LLM backbone, directly prompting
the LLM with the full catalog 
outperforms all baselines by at least 9.0 EM points. 
It also maintains single-digit latency
(8\,s), faster than or comparable to retrieval-based methods,
yielding a substantially better accuracy--latency trade-off.

\minihead{\sys's curation improves both search quality and efficiency across all backbones (RQ2)}
Across 9 frozen LLM backbones, \sys improves EM by 13.5 percentage points and FR by 10.1 points on average, with maximum EM gains reaching 20 points. These improvements are highly consistent, increasing both metrics across all 63 backbone--merchant combinations. The effect is particularly pronounced for \texttt{grok-4.1-fast-non-reasoning}, where \sys more than doubles EM from 12.8\% to 31.1\%. Crucially, these gains are specific to \sys{}'s learned curator rather than off-the-shelf context pruning. On \sharedbackbone, substituting Jina Reranker 3.5 or Gemini Embedding 2 at the same retention budget degrades accuracy on 5 and 3 of the 7 merchants respectively, whereas \sys{} improves all 7. 
Furthermore, despite the additional curation step, \sys reduces end-to-end latency for all nine LLMs by 28\% on average (and up to 71\%).

\minihead{Post-training provides substantial gains beyond curation (RQ3)}
Combining context curation with post-training nearly triples EM over the raw backbone (from 10.7\% to 31.0\%), yielding an additional 10.3 EM and 9.3 FR points beyond inference-time curation alone. Crucially, performance increases monotonically across all seven merchants at each stage of the pipeline (raw backbone $\rightarrow$ encoder-only $\rightarrow$ full curator $\rightarrow$ post-training), validating \sys's design by confirming that learned context organization and downstream post-training provide compounding, additive benefits.

\minihead{Case Study}
For \ref{q:wedding}, \sys retains the ground-truth product,
\textit{Cherry Flambé Matte Lip Whip} (15 g), at the 8th position of a compact context, increasing EM from $0\%$ to $100\%$ for \texttt{grok-4.1-fast-reasoning} and from $0\%$ to $70\%$ for \texttt{gemma-4-E4B-it}.
With the same curated context, \sys's post-training raises
\texttt{gemma}'s EM from $70\%$ to $100\%$.
Before post-training, \texttt{gemma} sometimes selects $23$~g
alternatives, describing them as ``close to the lightweight preference'', despite the
$15$~g ground-truth being available;
after post-training, it consistently selects ground-truth, better matching the user's soft preference of
\qpref{ideally under 10 grams}. We present a detailed failure analysis in Appendix~\ref{app:failure_cases}.

\begin{table*}[t]
\centering
\setlength{\tabcolsep}{5.0pt}
\setlength{\aboverulesep}{0pt}
\setlength{\belowrulesep}{0pt}
\renewcommand{\arraystretch}{1.25}
\footnotesize
\caption{Main results on seven SMBs. Merchant columns report exact
variant match (EM); the final block averages EM, feature reward (FR),
and end-to-end latency across merchants. Arrowed values show
\textcolor{SysGainBlue}{EM/FR gains} and \textcolor{SysGainGreen}{latency
reductions} over the same backbone with full-catalog prompting.}
\label{tab:main_results}
\resizebox{\textwidth}{!}{%
\begin{tabular}{@{}l *{7}{c} >{\columncolor{AvgEMTint}}c *{2}{c}@{}}
\toprule
\multirow{2}{*}{\textbf{Method / Model}}
& \multicolumn{7}{c}{\textbf{EM (\%) $\uparrow$}}
& \multicolumn{3}{c}{\textbf{Average}} \\
\cmidrule(lr){2-8} \cmidrule(r){9-11}
& BB & OON & HK & WCP & WB & FF & PF & EM (\%) $\uparrow$ & FR (\%) $\uparrow$ & Latency (s) $\downarrow$ \\
\midrule
\multicolumn{8}{l}{\textit{Main method comparison (LLM backbone: \sharedbackbone)}} & & & \\
RAG-Fusion & 18.8 & 0.1 & 1.8 & 15.0 & 11.3 & 2.0 & 16.0 & 9.3 & 48.5 & 6.4 \\
LongLLMLingua & 9.1 & 12.3 & 18.4 & 8.6 & 9.1 & 5.3 & 31.4 & 13.5 & 52.2 & 64.2 \\
TourRank & 6.1 & 13.0 & 16.7 & 9.0 & 10.1 & 14.3 & 26.2 & 13.6 & 54.6 & 56.7 \\
RAG [EmbeddingGemma 300M] & 28.8 & 23.8 & 17.0 & 16.7 & 26.1 & 7.3 & 45.7 & 23.6 & 62.9 & 5.7 \\
RAG [Gemini Embedding 2] & 37.3 & 18.4 & 10.9 & 32.3 & 31.8 & 12.9 & 44.6 & 26.9 & 62.3 & 3.4 \\
RAG [Jina Reranker 3.5] & 42.5 & 23.8 & 16.7 & 21.9 & 36.5 & 11.5 & 41.9 & 27.8 & 64.1 & 9.2 \\
\hdashline[1pt/1pt]
\sharedbackbone & 38.6 & 44.0 & 30.7 & 30.3 & 37.4 & 30.4 & 46.3 & 36.8 & 69.0 & 8.0 \\
\multicolumn{8}{l}{\textcolor{gray}{\textit{\quad Context selectors (25\% catalog retention)}}} & & & \\
\quad + EmbeddingGemma 300M & 34.5 & 39.3 & 25.5 & 20.9 & 44.9 & 26.0 & 45.7 & 33.8 & 67.9 & 7.1 \\
\quad + Gemini Embedding 2 & 41.5 & 37.2 & 29.2 & 32.2 & 45.8 & 34.2 & 40.4 & 37.2 & 70.1 & 6.8 \\
\quad + Jina Reranker 3.5 & 44.7 & 43.5 & 29.5 & 29.8 & 44.8 & 28.6 & 43.8 & 37.8 & 69.9 & 12.4 \\
\rowcolor{SysRowBlue}
+ \sys & 51.2 & 56.3 & 51.1 & 42.4 & 60.0 & 46.6 & 52.2 & \cellcolor{AvgEMTintSys}\textbf{51.4}\smbPB{+14.6} & 77.3\smbP{+8.3} & 5.1\smbPd{2.9} \\
\midrule
\multicolumn{8}{l}{\textit{\sys full pipeline evaluation}} & & & \\
gemma-4-E4B-it & 15.2 & 7.5 & 2.5 & 6.6 & 7.8 & 11.1 & 24.3 & 10.7 & 56.7 & 5.8 \\
\rowcolor{SysRowBlue}
+ \sys{} [encoder] & 25.2 & 9.6 & 10.3 & 12.5 & 14.8 & 13.2 & 28.9 & \cellcolor{AvgEMTintSys}\textbf{16.4}\smbPB{+5.6} & 60.6\smbP{+3.9} & 3.8\smbPd{2.0} \\
\rowcolor{SysRowBlue}
+ \sys{} [curator] & 38.2 & 11.7 & 12.9 & 13.1 & 17.9 & 14.7 & 36.2 & \cellcolor{AvgEMTintSys}\textbf{20.7}\smbPB{+10.0} & 65.1\smbP{+8.4} & 3.9\smbPd{1.9} \\
\rowcolor{SysRowBlue}
+ \sys{} [curator + post-train] & 49.3 & 25.1 & 19.2 & 22.8 & 36.2 & 21.3 & 43.3 & \cellcolor{AvgEMTintSys}\textbf{31.0}\smbPB{+20.3} & 74.4\smbP{+17.7} & 3.7\smbPd{2.1} \\
\midrule
\multicolumn{8}{l}{\textit{\sys context curation across other frozen LLM backbones}} & & & \\
grok-4.1-fast-reasoning & 40.6 & 44.0 & 25.7 & 18.1 & 29.8 & 21.6 & 28.1 & 29.7 & 65.4 & 33.8 \\
\rowcolor{SysRowBlue}
+ \sys & 58.9 & 50.6 & 46.8 & 39.2 & 61.1 & 39.6 & 53.6 & \cellcolor{AvgEMTintSys}\textbf{50.0}\smbPB{+20.3} & 78.2\smbP{+12.9} & 26.5\smbPd{7.3} \\
grok-4.1-fast-non-reasoning & 14.8 & 21.6 & 12.6 & 18.3 & 11.8 & 3.1 & 7.7 & 12.8 & 50.3 & 9.4 \\
\rowcolor{SysRowBlue}
+ \sys & 46.2 & 39.0 & 22.3 & 31.9 & 29.7 & 22.2 & 26.2 & \cellcolor{AvgEMTintSys}\textbf{31.1}\smbPB{+18.2} & 69.3\smbP{+18.9} & 2.7\smbPd{6.8} \\
kimi-k2-thinking & 40.1 & 39.4 & 21.9 & 22.4 & 34.6 & 23.3 & 32.0 & 30.5 & 65.8 & 70.7 \\
\rowcolor{SysRowBlue}
+ \sys & 57.1 & 49.5 & 33.9 & 37.9 & 54.5 & 38.4 & 47.1 & \cellcolor{AvgEMTintSys}\textbf{45.5}\smbPB{+15.0} & 75.8\smbP{+9.9} & 40.5\smbPd{30.2} \\
glm-5.2 & 32.6 & 28.8 & 15.8 & 31.2 & 42.4 & 33.1 & 47.0 & 33.0 & 60.2 & 122.7 \\
\rowcolor{SysRowBlue}
+ \sys & 54.1 & 43.4 & 30.5 & 42.5 & 56.0 & 43.7 & 59.2 & \cellcolor{AvgEMTintSys}\textbf{47.1}\smbPB{+14.1} & 73.9\smbP{+13.7} & 116.7\smbPd{6.0} \\
claude-sonnet-5 & 40.8 & 36.3 & 21.6 & 31.6 & 40.5 & 30.7 & 37.7 & 34.2 & 67.1 & 32.0 \\
\rowcolor{SysRowBlue}
+ \sys & 54.1 & 43.5 & 34.1 & 36.0 & 52.4 & 39.8 & 54.2 & \cellcolor{AvgEMTintSys}\textbf{44.9}\smbPB{+10.7} & 74.5\smbP{+7.4} & 27.3\smbPd{4.7} \\
gemini-3.1-pro & 57.5 & 54.7 & 48.8 & 49.1 & 51.2 & 51.5 & 52.5 & 52.2 & 77.0 & 106.7 \\
\rowcolor{SysRowBlue}
+ \sys & 69.9 & 70.8 & 68.7 & 55.7 & 52.3 & 55.5 & 65.0 & \cellcolor{AvgEMTintSys}\textbf{62.6}\smbPB{+10.4} & 83.3\smbP{+6.3} & 81.8\smbPd{24.9} \\
claude-opus-5 & 54.0 & 43.8 & 30.0 & 31.0 & 44.4 & 34.1 & 53.7 & 41.6 & 71.5 & 22.3 \\
\rowcolor{SysRowBlue}
+ \sys & 54.6 & 49.2 & 41.9 & 37.4 & 55.4 & 48.6 & 64.3 & \cellcolor{AvgEMTintSys}\textbf{50.2}\smbPB{+8.6} & 76.6\smbP{+5.1} & 21.1\smbPd{1.2} \\
\bottomrule
\end{tabular}%
}
\end{table*}

\subsection{Generalizability Test}
\label{subsec:generalizability}

\begin{figure}[t]
\centering
\begin{subfigure}[b]{0.61\textwidth}
  \centering
  \includegraphics[width=\textwidth]{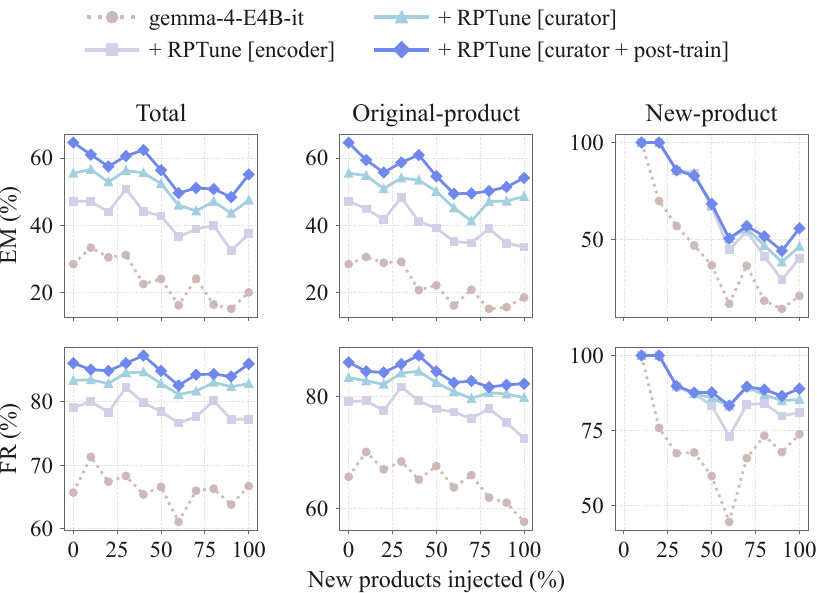}
  \caption{Generalizability to new products}
  \label{fig:catalog_increase}
\end{subfigure}\hfill
\begin{subfigure}[b]{0.37\textwidth}
  \centering
  \includegraphics[width=\textwidth]{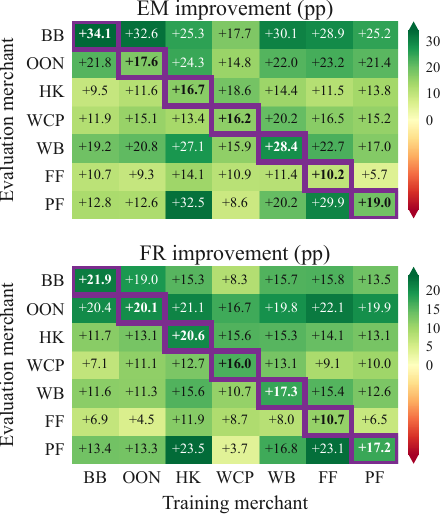}
  \caption{Generalizability to unseen merchants}
  \label{fig:cross_catalog}
\end{subfigure}
\caption{\sys generalizes well without retraining.
(a)~EM and FR on Beauty Bakerie as withheld products are injected back, for all
queries and for queries whose target is an original or a new
product. $0\%$ is the reduced catalog \sys was trained on and $100\%$ restores
the full catalog.
(b)~Gains over \texttt{gemma} for every train/eval
merchant pair, \textbf{all positive}; boxed diagonal cells are in-domain. Average out-of-domain gains are $+18.1$ EM and $+13.6$ FR, in-domain $+20.3$ and $+17.7$.}
\label{fig:generalization}
\end{figure}

For RQ4, we show that \sys generalizes well to
(1) newly added products within the same merchant
(Figure~\ref{fig:catalog_increase}) and
(2) unseen merchants (Figure~\ref{fig:cross_catalog}).

\minihead{\sys generalizes to new products without retraining}
Merchants frequently update inventory, making constant retraining impractical. To evaluate zero-shot generalization, we train \sys on a reduced Beauty Bakerie catalog, formed by randomly removing half of the products within each product type. For evaluation queries whose original ground-truth product was removed, we re-establish the new best-matching variant by re-running our pipeline's self-consistency filtering and human verification steps on the reduced catalog. We then measure performance as the withheld products are incrementally injected back (Figure~\ref{fig:catalog_increase}) with the same 100 queries evaluated at every injection level (details in Appendix~\ref{app:catalog_increase}). \sys proves highly resilient: as the catalog doubles, post-trained \sys experiences only a 14.7\% relative EM decline, half of the baseline's 29.5\% drop, with FR remaining practically unchanged. 
Crucially, \textit{\sys consistently outperforms the baseline at every injection level} for queries targeting both original and new products. Even at double the initial catalog size, \sys maintains absolute gains of 35.0 EM and 19.2 FR over the baseline.

\minihead{\sys generalizes to unseen merchants}
To evaluate zero-shot transfer across domains, we cross-evaluate each of our seven merchant-specific \sys models on all available catalogs (Figure~\ref{fig:cross_catalog}). The results are uniformly positive: \textit{all $7\times7$ combinations improve both EM and FR over the \texttt{gemma} baseline}. 
These robust improvements persist 
despite severe lexical shifts across catalogs with mean pairwise Jaccard similarity 
of 0.146.

\subsection{Ablation Analysis}
\label{subsec:ablation_analysis}

\begin{figure}[t]
\centering
\begin{minipage}[t]{0.40\textwidth}
  \centering
  \includegraphics[width=\linewidth]{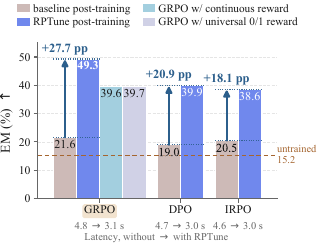}
  \caption{\sys improves \texttt{gemma}'s EM for all post-training objectives with low latency. \sys uses \dflt{GRPO}.
  }
  \label{fig:posttrain}
\end{minipage}\hfill
\begin{minipage}[t]{0.545\textwidth}
  \centering
  \includegraphics[width=\linewidth]{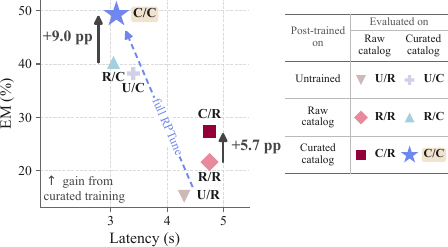}
  \caption{
  EM vs.\ latency for every post-train/eval combination; \dflt{C/C} is \sys.
  Post-training on curated contexts consistently improves search accuracy.
  }
  \label{fig:ablation_pareto_curation}
\end{minipage}
\end{figure}

\begin{figure}[t]
\centering
\begin{minipage}[t]{0.60\textwidth}
  \vspace{0pt}
  \centering
  \includegraphics[width=0.605\linewidth]{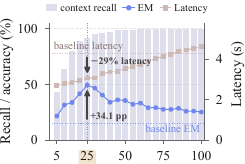}\hspace{2pt}%
  \includegraphics[width=0.385\linewidth]{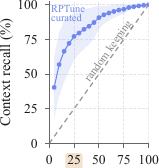}\\[-4pt]
  {\scriptsize Retention budget (\% of catalog kept)}
  \caption{(Left) Pruning sweep on Beauty Bakerie. (Right) Context recall over
  7 merchants; shading is 1 STD across merchants. \sys uses \dflt{25\%} as the default budget.}
  \label{fig:pruning_sweep}
\end{minipage}\hfill
\begin{minipage}[t]{0.37\textwidth}
  \vspace{0pt}
  \centering
  \includegraphics[width=0.98\linewidth]{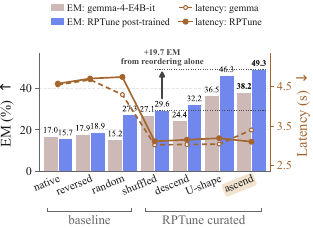}
  \caption{EM vs.\ latency for catalog ordering variants on Beauty
  Bakerie; \sys uses \dflt{ascend} as default.
  }
  \label{fig:ordering}
\end{minipage}
\end{figure}

We ablate five design choices in \sys: (1) post-training objective (Figure~\ref{fig:posttrain}); (2) use of curated contexts during post-training (Figure~\ref{fig:ablation_pareto_curation}); (3) pruning budget (Figure~\ref{fig:pruning_sweep}); (4) context ordering (Figure~\ref{fig:ordering}); and
(5) reorganizer training objective. Supplementary FR results in Appendix~\ref{app:fr_views} are consistent with all findings in this section.

\minihead{\sys improves performance across post-training objectives}
We compare GRPO with DPO and IRPO on synthetic data, with and without \sys curation (Figure~\ref{fig:posttrain}), and additionally train GRPO on \sys-curated contexts with two alternatives to our context-relative reward: (i)~a \emph{continuous} reward, $R = y(q, c_{\mathrm{pred}})$, and (ii)~a \emph{global} top-1 reward that grants $R = 1$ only when $c_{\mathrm{pred}}$ is the best variant in the full catalog rather than in $\widetilde{\mathcal{C}}_q$.
Post-training alone consistently improves the \texttt{gemma} baseline across all methods, validating our data pipeline. 
\sys curation further enhances all methods (averaging +22.2 pp EM and a 34\% latency reduction). 
Of all, GRPO achieves the strongest overall performance and benefits most from \sys.
\sys's context-relative reward is essential: EM on BB drops 10 points with both continuous and global rewards.

\minihead{\sys-curated contexts improve post-training}
We isolate the impact of context curation during post-training by training \texttt{gemma-4-E4B-it} on either raw (R) or curated (C) catalogs and evaluating on both (Figure~\ref{fig:ablation_pareto_curation}). Beyond accelerating training via shorter rollouts, training on curated contexts strictly improves downstream model quality across all evaluation settings. Specifically, curated training improves EM by 5.7 points on raw evaluation contexts (C/R vs. R/R) and 9.0 points on curated evaluation contexts (C/C vs. R/C). Ultimately, applying curation at both training and inference (C/C) maximizes EM and cuts latency by 35\% versus the raw baseline (R/R).

\minihead{\sys maintains gains across pruning budgets}
We sweep the context retention budget from 5\% to 100\% for \texttt{gemma} with full \sys (Figure~\ref{fig:pruning_sweep}). First, \sys outperforms the raw full-catalog baseline at every tested budget, including full retention. Second, our curation achieves high recall with small contexts: \sys's default 25\% retention rate yields 77.1\% average context recall (over 3$\times$ better than random selection), while 50\% retention reaches 90.7\%. Finally, downstream accuracy peaks at an intermediate budget: on Beauty Bakerie, EM and FR peak at 25\% retention (with 92\% recall). 
Larger budgets add marginal recall but lower accuracy and raise latency: \sys's pruning effectively filters out noise for this sweet spot.

\minihead{\sys improves performance across context orderings}
We evaluate the raw and \sys-post-trained \texttt{gemma} on Beauty Bakerie under different context layouts (Figure~\ref{fig:ordering}). Baseline contexts use \textit{(1)} random, \textit{(2)} native, or \textit{(3)} reversed native ordering of the full catalog, whereas, based on context priority scores, \sys-curated contexts use \textit{(4)} ascending, \textit{(5)} descending, \textit{(6)} U-shaped (most relevant at both ends), or \textit{(7)} randomly shuffled ordering on the 25\% retained items. 
For both models, every \sys curated layout outperforms all baseline layouts, gaining 16.8 EM points while cutting latency by 32\%, where the ascending order performs best, followed by U-shaped. 

\minihead{\sys's LLM-in-the-loop reorganizer objective optimizes contexts for LLM interpretation}
To isolate the value of LLM feedback, we replace the reorganizer objective in Eq.~\ref{eq:reorganizer_loss} with a supervised listwise cross-entropy that fits $w_\theta$ directly to the relevance labels in $\mathcal{D}_{\text{rel}}$ (Appendix~\ref{app:supervised_reorganizer}). With frozen \texttt{gemma}, the supervised reorganizer reaches 16.4\% average EM, no better than the encoder alone and 4.3 points below \sys's reorganizer, which achieves higher EM on all 7 merchants.

\section{Related Work}
\label{sec:relatedwork}

\minihead{Retrieval and context construction}
Standard retrieval-augmented generation (RAG) pipelines
retrieve candidates using lexical methods \citep{robertson2009bm25}, dense encoders \citep{geminiembedding2, embeddinggemma2025, zhang2025qwen3embedding}, or rankers \citep{zhang2025qwen3embedding, nasika2026jinarerankerv35efficientlistwisereranker}.
More elaborate frameworks combine LLMs with multi-query
retrieval and rank fusion~\citep{Rackauckas_2024} or
tournament-based ranking~\citep{chen2025tourrankutilizinglargelanguage}.
Beyond candidate retrieval, query-aware compression and
context reordering improve the information density of
LLM inputs~\citep{jiang-etal-2023-llmlingua,jiang2024longllmlinguaacceleratingenhancingllms}.
These methods provide practical solutions for large
marketplaces; however,
we show that in-context product search is more promising for SMBs.

\minihead{LLM post-training}
LLM post-training adapts models using preference supervision
or task-specific rewards.
DPO~\citep{rafailov2024directpreferenceoptimizationlanguage} learns from preferred and
dispreferred responses, which IRPO~\citep{wu2025incontextrankingpreferenceoptimization}
extends to ranked candidate lists.
Reinforcement learning methods \citep{shao2024deepseekmath, yu2025dapoopensourcellmreinforcement} have improved LLM performance further on
complex reasoning tasks.
Building on these advances, we post-train LLMs for in-context product search on curated catalogs.

\section{Conclusion}
\label{sec:conclusion}

We introduced \sys, an in-context catalog search framework tailored for SMBs. Trained entirely via synthetic supervision, \sys uses downstream LLM feedback to curate compact, position-aware catalog contexts, which subsequently enhance LLM post-training for product selection. 
Across 700 complex conversational queries from 7 real-world merchants, \sys significantly improves search accuracy for all 9 evaluated proprietary and open-weight LLMs.

\bibliography{main}
\bibliographystyle{abbrvnat}

\appendix
\newpage
\section{Appendix}

\subsection{Training Data Generation}
\label{app:training_data}
We generate synthetic query--product pairs $\mathcal{D}_{\text{pair}}$ and catalog-wide
relevance scores $\mathcal{D}_{\text{rel}}$ in two steps.

\textit{Query--Product Pair Generation.} For each product variant $c_i$ in the merchant's catalog, we prompt a frontier LLM (\texttt{gemini-3.1-pro}) to generate 10 diverse search queries that intentionally target $c_i$. To ensure diversity and realism, the generation prompt enforces that the queries span various user intents (e.g., shade-matching, finish-specific, and/or occasion-based)
and combine multiple product features (e.g., price, taxonomy tags, size, and/or
product type) into a short, natural 1-2 sentence search query. This yields $\mathcal{D}_{\text{pair}} = \{(q, c_{q})\}$, a set of
$10n$ realistic queries, each paired with its target variant $c_{q}$.
For example, one query generated for the ground-truth variant of \ref{q:wedding} in
\textit{Beauty Bakerie} (\texttt{Cherry Flambé Matte Lip Whip}) is \textit{(Q2)}\texttt{``fiery red liquid matte lipstick for date night''}.

\textit{Catalog-Wide Relevance Scoring.} A single positive per query is
too coarse a signal: retrieval training also needs to know which variants
are partial matches or near-substitutes. For each query $q$, we therefore
prompt \texttt{gemini-3.1-pro} as an LLM relevance judge to assign a graded score
$y(q, c) \in [0.0, 1.0]$ to every variant $c$ in the catalog. Scoring runs
in batches of 10 variants per call, and each call additionally
receives a running dictionary of the scores already assigned for $q$; the
judge may revise earlier scores so that the final ranking is globally
consistent across the catalog. The result is $\mathcal{D}_{\text{rel}} = \{(q, c, y(q,c))\}$, a dense
relevance distribution over the entire catalog for every query.
Note that the target product $c_q$ used during query generation does not necessarily receive a score of $1.0$.

This relevance-scoring procedure is distinct from in-context
catalog search: it assigns graded scores to query--product
pairs rather than selecting a single best-matching product
for a user query.
Two design choices simplify the labeling task.
First, each call evaluates only 10 product records, 
rather than processing all product descriptions at once.
Second, the synthetic queries are short and simple, in
contrast to the complex, multi-constraint real-world conversational
queries (e.g., \ref{q:wedding}).

\subsection{Training Curves}
\label{app:curves}

\begin{figure*}[htbp]
    \centering
    \includegraphics[width=0.5\textwidth]{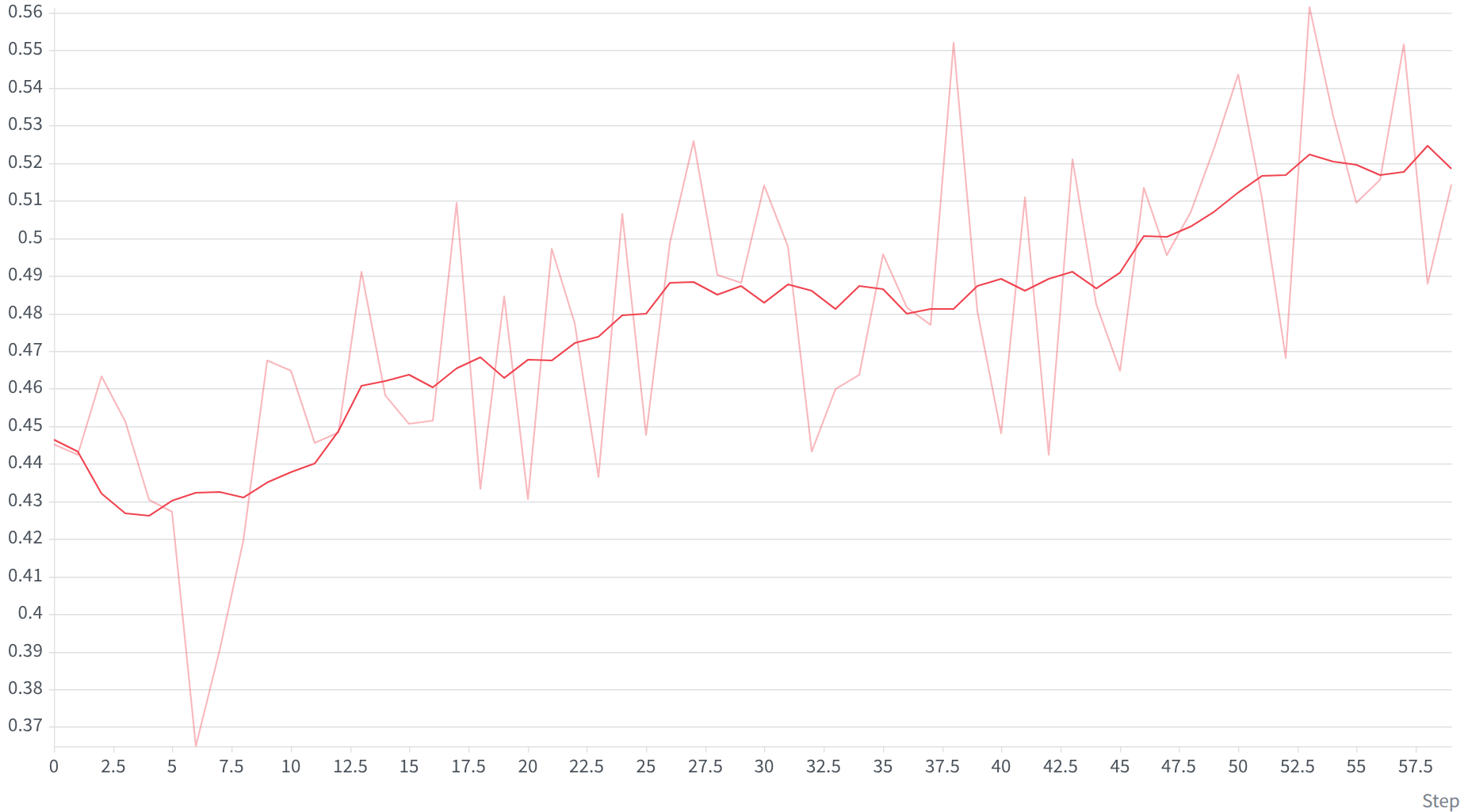}
    \caption{Average relevance score of the products selected by the frozen LLM during \sys reorganizer training.}
    \label{fig:reward_curve}
\end{figure*}

We use $G=16$ rollouts per query and minibatches of $B=128$ queries for reorganizer training. Figure~\ref{fig:reward_curve} reports the average relevance score of the products selected by the frozen LLM during training, which increases steadily over the course of training. Every optimizer update contains informative query groups with non-zero advantages, with at least 44 such groups per update. The invalid-output (where $r_g=-1$) rate is only $0.83\%$, while cases in which different selected products receive identical relevance scores account for $6.2\%$ of query groups; thus, the training signal is well-defined and informative.

\subsection{Evaluation Data}
\label{app:data}
We evaluate \sys and baselines on 7 real-world SMBs spanning
distinct retail verticals, including pet supplies, sustainable
lifestyle, and eyewear.

\minihead{Catalog collection}
All seven storefronts run on Shopify, which exposes the full product listing at
a public endpoint: appending \texttt{/products.json} to the storefront URL
returns every product together with its variants, options, prices, images and
descriptions. The entire catalog is therefore obtainable in a single request per
merchant, with no scraping, credentials or merchant cooperation
required.\footnote{The endpoint pages at 250 products per request, so the one
catalog above that size (WB, 277 products) is fetched with
\texttt{?limit=250\&page=$n$}.} This is precisely the regime \sys targets: an
SMB can hand its whole catalog to a long-context model without any retrieval
infrastructure. The storefronts are Beauty Bakerie
(\url{https://www.beautybakerie.com/}), Oats Overnight
(\url{https://oatsovernight.com/}), The Honest Kitchen
(\url{https://www.thehonestkitchen.com/}), Wilsun Custom Products
(\url{https://wilsuncustomproducts.com/}), Wairua Beauty
(\url{https://wairuabeauty.com/}), Focus \& Frame Eyewear
(\url{https://focusandframeeyewear.com/}) and Package Free
(\url{https://packagefreeshop.com/}).

To avoid relying on sensitive
real-user search logs, we design a three-stage pipeline to
synthesize conversational queries that reflect realistic user
intents without trivially matching individual product descriptions following prior work on LLM-generated evaluation data with human
quality checks \citep{yao2024taubenchbenchmarktoolagentuserinteraction, chen2023benchmarkinglargelanguagemodels, qin2023toolllmfacilitatinglargelanguage}.
The resulting evaluation set contains 100 queries per merchant.

Stage \stepnum{1}: \textit{Query Generation.}
We instruct a multimodal LLM (\texttt{gemini-3.1-pro}) to act as a synthetic customer by
providing it with high-level context: (1) general merchant
background, (2) a full-page screenshot of the merchant's
``About Us'' webpage capturing brand identity and aesthetics,
and (3) aggregate catalog feature distributions (e.g.,
product types, tags, prices, and weights). Importantly,
we withhold individual product records to discourage queries
tailored to specific product descriptions. We prompt the
model to generate a single natural-sounding and realistic conversational
query with multiple complex constraints to encourage reasoning
about trade-offs rather than simple keyword matching.

Stage \stepnum{2}: \textit{Query Refinement.}
We provide each candidate query and the complete catalog to the same model and collect 10 recommendations, each using a
different randomly shuffled catalog order. In each trial,
the model selects the single product variant that best
satisfies the query's constraints.

We define the
\emph{consensus score} as the maximum number of trials
selecting the same variant.
Because catalog order is randomized across trials, a query with one clearly
best variant is resolved consistently regardless of order, whereas a query
with several near-equivalent candidates yields selections that vary with
order. We therefore use consensus as a heuristic for query difficulty (see
\emph{Evaluation robustness to consensus-based filtering} in the evaluation
characteristics below):
scores $\geq 8$ indicate highly stable selections characteristic of easy queries, whereas scores $\leq 3$ indicate unstable selections, reflecting potential ambiguity or overly restrictive constraints.
We retain queries with moderate consensus (scores between $4$ and $7$, inclusive). For queries with a unique maximum, we assign the variant that achieved this consensus as the reference label. In cases where multiple variants tie for the maximum score (which occurs in 10.9\% of the retained queries), we preserve the set of tied variants as candidate labels to be resolved in Stage~\stepnum{3}.
For queries outside this range, we return the query and its consensus score to the generator in Stage~\stepnum{1}, instructing it to introduce more complex trade-offs when the score is too high and to relax or clarify constraints when it is too low.
This refinement loop converges in ${\sim}3$ rounds on average.

Stage \stepnum{3}: \textit{Human Verification.}
For each merchant, we first generate 10 pilot queries using
Stages~\stepnum{1} and~\stepnum{2}. An expert assesses whether
each query is natural-sounding and realistic, then selects
the best-matching product variant without seeing the reference
label(s). We compare the expert's selections with the reference
labels and revise the generation prompt to address any issues
identified during review. 
Large-scale generation begins only
after all 10 pilot queries pass expert review.

After large-scale generation, we manually audit a 10\% subset of standard queries and all tied cases using the same blinded selection procedure. For tied cases, if the expert's selection matches a variant within the tie set, that specific variant is finalized as the single ground-truth label. Any disagreements are reviewed by a panel of 6 experts, and pairs whose reference labels are not supported by full expert consensus are discarded. The overall rejection rate across this auditing process is 9.6\%.

\minihead{Evaluation characteristics}
We discuss the following four aspects of the evaluation data.

\underline{\textit{Query-level distribution shift.}}
The generated evaluation queries represent out-of-distribution data relative to the training data described in Section \ref{subsec:training}. While the training queries are typically short, concise, and easily resolved by standard LLMs, our evaluation queries
combine multiple interacting constraints, implicit preferences,
and trade-offs in natural conversational language. 
\ref{q:wedding} is a representative evaluation query for
\textit{Beauty Bakerie} that introduces conflicting constraints (e.g., wedding-grade performance vs.\ a sub-\$25 budget) and implicit requirements (e.g., compact clutch-sized packaging), establishing a highly realistic and challenging retrieval scenario.

\underline{\textit{Evaluation scope.}}
We focus on \emph{single-turn} queries and leave multi-turn
evaluation to future work. Nevertheless, these queries
consolidate preferences and constraints that could emerge
across multiple turns of a shopping conversation. For example, \ref{q:wedding} could be decomposed into successive turns
specifying shade and finish, occasion and wear context, ethical
preferences, product weight, and budget. \sys can be easily adapted to this setting by running
the encoder--reorganizer at each turn, conditioned on the
accumulated constraints.

\underline{\textit{Validity of model-assigned labels.}}
Because \texttt{gemini-3.1-pro} assigns the reference labels, they may
favor Gemini-family backbones. We address this risk in three ways.
First, Stage~\stepnum{3} includes blinded human verification of the
pilot queries, all tied cases, and a 10\% sample of the remaining
queries, which rejects 9.6\% of query--label pairs overall. Second,
\sys's gains are not specific to the labeler's family: in
Table~\ref{tab:main_results}, the average gain for Grok (+19.3 EM)
exceeds that for Gemini (+12.5 EM). Third, we audit the labels with a
panel of independent LLM judges.

For the audit, we sample 154 evaluation queries, stratified by merchant
(22 per merchant). Each query's candidate pool (11.1 candidates on
average) draws on four sources: (i)~the reference label and every
variant selected in the labeler's 10 Stage~\stepnum{2} runs;
(ii)~\texttt{claude-opus-5}'s modal selection over 10 shuffled
full-catalog runs, i.e., Stage~\stepnum{2} with a different labeler;
(iii)~the selections of the evaluated backbones for which we retain
per-run predictions (\texttt{glm-5.2}, \texttt{kimi-k2-thinking}, both
Grok backbones, and \texttt{gemini-3.7-flash} with Gemini Embedding 2
selection for every merchant, plus \texttt{gemini-3.1-pro},
\texttt{gemini-3.7-flash}, \texttt{claude-opus-5}, and
\texttt{claude-sonnet-5}); and (iv)~five distractors: the three
top-ranked BM25 results not already in the pool and two random catalog
items. Four non-Gemini judges (\texttt{claude-opus-5},
\texttt{gpt-oss-120b}, \texttt{qwen3-235b-a22b-instruct-2507}, and
\texttt{deepseek-r1-0528}) see each pool in random order with sources
hidden. They grade every candidate from 0 to 3 (0: wrong product;
1: right kind of product but misses a stated priority or firm
requirement; 2: a defensible recommendation that satisfies the
customer's top priorities; 3: meets every stated requirement) and name
the single best variant. Each judge makes three passes over reshuffled
pools, yielding 12 judgments per candidate (4 judges $\times$ 3
passes). A variant is \emph{acceptable} if at least 6 of these
judgments grade it $\geq 2$.

The labels are informative rather than arbitrary. The panel accepts
the reference label for 64.9\% of queries, compared with 1.1\% for
BM25 distractors and 0\% for random ones. It accepts
\texttt{claude-opus-5}'s full-catalog selection for only 55.8\% of queries, even though \texttt{claude-opus-5} is itself a judge
and may favor its own choice, whereas no judge shares the labeler's
model family. A single reference nonetheless cannot capture every
acceptable answer, because many queries admit several defensible
choices. On average, 1.85 variants per query are acceptable
(44.8\% of queries have at least two), and
on identical pools, pairwise agreement between judges on the best
variant ranges from only 36.4\% to 46.1\%. \texttt{claude-opus-5}'s
full-catalog selection matches the reference label for 44.2\% of
queries. That falls within the inter-judge range, so the labeler and
an independent model disagree no more often than the judges disagree
with each other. We therefore do not interpret
\texttt{gemini-3.1-pro}'s absolute scores in
Table~\ref{tab:main_results} as evidence that it is the strongest
backbone. Our conclusions about \sys instead
emphasize within-backbone improvements observed across both
Gemini and non-Gemini models.

\underline{\textit{Evaluation robustness to consensus-based filtering.}}
Because Stage~\stepnum{2} retains only moderate-consensus queries, our
evaluation set could favor methods that reduce sensitivity to catalog order.
To assess whether \sys's benefits extend beyond these queries, we set aside
100 Beauty Bakerie candidates that received consensus scores $\geq 8$ during
Stage~\stepnum{2} (that have been returned to the generator
for revision), and use them as a test set. We label each with its majority
variant and evaluate both raw \texttt{gemma-4-E4B-it} and the full \sys
pipeline. As expected, these
queries are considerably easier: raw Gemma reaches 33.9 EM and 72.0 FR,
compared with 15.2 EM and 64.5 FR on Beauty Bakerie's main evaluation set (Table \ref{tab:full_results}).
\sys nonetheless delivers large gains, reaching 87.8 EM and 94.6 FR and
eliminating roughly 80\% of the baseline's remaining errors on both metrics
(EM error falls from 66.1 to 12.2, and FR error from 28.0 to 5.4). The higher baseline performance on these high-consensus queries supports
using consensus as a difficulty-control heuristic, and \sys's large gains on
them indicate that consensus filtering does not bias the evaluation set
toward methods that reduce catalog-order sensitivity.

\subsection{Models and Baselines}
\label{app:setups}
We evaluate \sys across 9 LLM backbones from multiple model families, including \texttt{gemini-3.1-pro}~\citep{gemini31pro}, \texttt{gemini-3.7-flash}~\citep{gemini37flash}, \texttt{claude-sonnet-5}~\citep{claude_sonnet5}, \texttt{claude-opus-5}~\citep{claude_opus5}, \texttt{grok-4.1-fast-reasoning} and \texttt{grok-4.1-fast-non-reasoning}~\citep{grok41fast}, \texttt{kimi-k2-thinking}~\citep{kimi_k2_thinking}, \texttt{glm-5.2}~\citep{glm52}, and \texttt{gemma-4-E4B-it}~\citep{gemma4}. 
For each backbone, we compare raw full-catalog prompting, with a randomly shuffled catalog in each trial, against \sys-curated contexts retaining the top-ranked 25\% of catalog variants. 
To further isolate curation quality, we compare the \sys curator against three alternative context selectors: two dense retrieval models (EmbeddingGemma 300M~\citep{embeddinggemma2025}, upon which the \sys encoder is trained, and Gemini Embedding 2~\citep{geminiembedding2}) and a learned listwise ranker (Jina Reranker 3.5~\citep{nasika2026jinarerankerv35efficientlistwisereranker}). All baselines retain the top-ranked 25\% of catalog variants to match \sys's retention budget.
These comparisons keep the downstream LLM weights frozen to isolate the effect of context curation. We evaluate \sys's post-training module with \texttt{gemma-4-E4B-it}
post-trained on curated contexts.

We compare \sys with retrieval, ranking, and context-compression
baselines, using \texttt{gemini-3.7-flash} as the shared backbone
for their LLM-based configurations.
We evaluate Gemini Embedding 2 and Jina Reranker 3.5 within a RAG pipeline where the LLM selects from the top 10 variants.
We additionally compare with
RAG-Fusion~\citep{Rackauckas_2024,medrano2026scalingretrievalaugmentedgeneration},
which uses Gemini Embedding 2 to retrieve candidates for
multiple generated queries and combines their rankings
using reciprocal rank fusion;
TourRank~\citep{chen2025tourrankutilizinglargelanguage},
which aggregates results from repeated LLM-based tournaments
to rank candidates; and
LongLLMLingua~\citep{jiang2024longllmlinguaacceleratingenhancingllms},
which performs query-aware context compression.

We additionally compare
GRPO with two alternative post-training objectives in Section~\ref{subsec:ablation_analysis}:
DPO~\citep{rafailov2024directpreferenceoptimizationlanguage}
and IRPO~\citep{wu2025incontextrankingpreferenceoptimization},
which extends preference optimization to ranking tasks.
All three use the same synthetic catalog-grounded training
data and backbone. We train GRPO for one epoch and DPO and
IRPO for three epochs.

\minihead{Setups}
Finally, we detail our inference and training infrastructure. For proprietary models (Gemini, including Gemini Embedding 2; Claude; Grok) we call their respective APIs, while \texttt{kimi-k2-thinking} and \texttt{glm-5.2} are evaluated via their Model-as-a-Service (MaaS) endpoints on Google Cloud Vertex AI Model Garden. For custom and open-weights models, we utilize H100 GPUs: \texttt{gemma-4-E4B-it} is trained on two 80GB H100 GPUs using Unsloth \citep{unsloth} and evaluated on a single 80GB H100 GPU. Similarly, we evaluate Jina Reranker 3.5, and train and evaluate the \sys{} curator, on a single 80GB H100 GPU.

\subsection{Feature Reward Definitions}
\label{app:metrics}

Let $\hat v$ be the variant selected by the model and $v^\star$ the ground-truth variant, with parent products $\hat p$ and $p^\star$.
The feature reward (FR) is
\begin{equation}
\mathrm{FR}(\hat v, v^\star) =
\begin{cases}
0 & \text{if } \hat v \notin \mathcal{V} \text{ (invalid ID, parse error, or failed call)},\\
1 & \text{if } \hat v = v^\star,\\
\dfrac{1}{Z}\sum_{f\in\mathcal{F}} w_f\, r_f(\hat v, v^\star) & \text{otherwise,}
\end{cases}
\end{equation}
where $\mathcal{V}$ is the set of catalog variants and $Z=\sum_f w_f = 4.8$.
The eight feature scores $r_f \in [0,1]$ and their weights are:

\paragraph{Categorical attributes.}
Each is an indicator of equality. We define $r_\text{type}$ as:
$$r_{\text{type}} = \mathbb{1}[\hat p.\texttt{product\_type} = p^\star.\texttt{product\_type}],$$ with $w=0.5$;
$r_{\text{ship}}$, $r_{\text{tax}}$, $r_{\text{avail}}$ are the same indicator on the variant fields
\texttt{requires\_shipping}, \texttt{taxable}, and \texttt{available} ($w=0.1$ each).

\paragraph{Numeric attributes.}
For $x \in \{\texttt{price}, \texttt{grams}\}$ ($w=1.0$ each), with relative difference $\delta_x$ defined as:
$$\delta_x = |\hat v.x - v^\star.x| / v^\star.x,$$
we define $r_x$ as: 
\begin{equation}
r_x =
\begin{cases}
1 & \delta_x \le 0.1,\\
\max(0,\, 1-\delta_x) & \text{otherwise,}
\end{cases}
\end{equation}
and if $v^\star.x = 0$, then $r_x = \mathbb{1}[\hat v.x = 0]$.

\paragraph{Tags.}
With tag sets $T(\cdot)$ ($w=1.0$), we define $r_{\text{tag}}$ as:
\begin{equation}
r_{\text{tag}} =
\begin{cases}
\mathbb{1}[T(\hat p)=\emptyset] & T(p^\star)=\emptyset,\\
|T(\hat p)\cap T(p^\star)| \,/\, |T(p^\star)| & \text{otherwise.}
\end{cases}
\end{equation}

\paragraph{Description.}
With $B(\cdot)$ the set of lower-cased word bigrams of \texttt{body\_html} ($w=1.0$),
$r_{\text{body}} = |B(\hat p)\cap B(p^\star)| \,/\, |B(p^\star)|$.
If the target has no bigrams, the same ratio is computed over unigrams;
if it has no words at all, then $r_{\text{body}}=\mathbb{1}[\hat p \text{ has no words}]$.

All overlap scores are recall-oriented: they are normalized by the target's features, so the selected product is not penalized for having extra tags or text.

\paragraph{Example.}
On Beauty Bakerie, suppose the target is \emph{SophistiCAKED Deluxe Powder Brush} (\$14.00, 62\,g) and the model selects \emph{Milk \& Honey Highlighting Brush} (\$18.00, 30\,g).
Both are \emph{Makeup Essentials} and match on shipping, taxable, and available,
so the categorical terms contribute $0.5+0.1+0.1+0.1=0.8$.
Price: $\delta=|18-14|/14=0.286>0.1$, so $r_{\text{price}}=0.714$.
Grams: $\delta=|30-62|/62=0.516$, so $r_{\text{grams}}=0.484$.
All 3 target tags (\emph{back-in-stock}, \emph{essentials}, \emph{face product}) appear in the selected product, so $r_{\text{tag}}=1$.
Only 2 of the target's 43 description bigrams overlap, so $r_{\text{body}}=0.047$.
Hence
\[
\mathrm{FR} = \frac{0.8 + 0.714 + 0.484 + 1 + 0.047}{4.8} = \frac{3.045}{4.8} = 0.634,
\]
whereas exact match (EM) gives $0$ for this prediction.

\subsection{Full Merchant-wise Results}
\label{app:full_results}

Table~\ref{tab:main_results} reports per-merchant exact match together with
averaged EM, feature reward and latency. Table~\ref{tab:full_results} gives the
underlying breakdown: all three metrics for every merchant, in the same row
order. Averaging a block across the seven merchants reproduces the
corresponding Average column of Table~\ref{tab:main_results}. The only exception
is the dashed row in the \texttt{gemma-4-E4B-it} block, the supervised-reorganizer
ablation of Section~\ref{subsec:ablation_analysis}
(Appendix~\ref{app:supervised_reorganizer}), which has no counterpart in
Table~\ref{tab:main_results}.

\begin{table*}[t]
\centering
\setlength{\tabcolsep}{2pt}
\renewcommand{\arraystretch}{1.05}
\footnotesize
\caption{Merchant-wise breakdown of Table~\ref{tab:main_results}: exact variant match
(EM, \%, $\uparrow$), feature reward (FR, \%, $\uparrow$) and end-to-end latency
(Lat., s, $\downarrow$) per merchant, in the same row order as
Table~\ref{tab:main_results}. The EM columns are exactly the per-merchant
values of Table~\ref{tab:main_results}; averaging a row across the seven merchants
reproduces the corresponding Average column. The dashed row in the
\texttt{gemma-4-E4B-it} block is an ablation without a counterpart in
Table~\ref{tab:main_results}: the curator's reorganizer is trained with the supervised
objective of Appendix~\ref{app:supervised_reorganizer} instead of LLM-in-the-loop RL
(Section~\ref{subsec:ablation_analysis}).}
\label{tab:full_results}
\begin{adjustbox}{max width=\textwidth}
\begin{tabular}{l ccc ccc ccc ccc ccc ccc ccc}
\toprule
\multirow{2}{*}{\textbf{Method / Model}}
& \multicolumn{3}{c}{\textbf{BB}}
& \multicolumn{3}{c}{\textbf{OON}}
& \multicolumn{3}{c}{\textbf{HK}}
& \multicolumn{3}{c}{\textbf{WCP}}
& \multicolumn{3}{c}{\textbf{WB}}
& \multicolumn{3}{c}{\textbf{FF}}
& \multicolumn{3}{c}{\textbf{PF}} \\
\cmidrule(lr){2-4} \cmidrule(lr){5-7} \cmidrule(lr){8-10} \cmidrule(lr){11-13} \cmidrule(lr){14-16} \cmidrule(lr){17-19} \cmidrule(lr){20-22}
& EM & FR & Lat. & EM & FR & Lat. & EM & FR & Lat. & EM & FR & Lat. & EM & FR & Lat. & EM & FR & Lat. & EM & FR & Lat. \\
\midrule \multicolumn{22}{l}{\textit{Main method comparison (LLM backbone: \sharedbackbone)}} \\
\addlinespace[1pt]
RAG-Fusion & 18.8 & 62.3 & 6.0 & 0.1 & 25.7 & 6.4 & 1.8 & 39.3 & 7.0 & 15.0 & 55.3 & 5.8 & 11.3 & 55.4 & 6.5 & 2.0 & 55.5 & 6.7 & 16.0 & 46.1 & 6.7 \\
LongLLMLingua & 9.1 & 45.0 & 26.5 & 12.3 & 49.6 & 25.4 & 18.4 & 54.6 & 63.4 & 8.6 & 44.0 & 65.5 & 9.1 & 56.8 & 83.2 & 5.3 & 55.8 & 143.6 & 31.4 & 59.9 & 42.0 \\
TourRank & 6.1 & 41.2 & 58.2 & 13.0 & 50.8 & 55.0 & 16.7 & 54.8 & 60.6 & 9.0 & 50.5 & 58.2 & 10.1 & 58.3 & 59.1 & 14.3 & 65.2 & 54.4 & 26.2 & 61.5 & 51.4 \\
RAG [EmbeddingGemma 300M] & 28.8 & 66.9 & 5.2 & 23.8 & 61.0 & 6.1 & 17.0 & 49.4 & 6.8 & 16.7 & 59.4 & 5.8 & 26.1 & 63.9 & 4.7 & 7.3 & 68.5 & 5.5 & 45.7 & 71.3 & 5.8 \\
RAG [Gemini Embedding 2] & 37.3 & 73.9 & 3.6 & 18.4 & 51.4 & 3.2 & 10.9 & 47.5 & 3.8 & 32.3 & 63.7 & 3.5 & 31.8 & 68.8 & 3.0 & 12.9 & 59.0 & 3.1 & 44.6 & 71.5 & 3.3 \\
RAG [Jina Reranker 3.5] & 42.5 & 74.4 & 4.5 & 23.8 & 58.9 & 4.3 & 16.7 & 50.5 & 5.2 & 21.9 & 61.5 & 10.2 & 36.5 & 69.0 & 7.2 & 11.5 & 62.9 & 12.8 & 41.9 & 71.4 & 20.1 \\
\hdashline[1pt/1pt]
\addlinespace[1pt]
\sharedbackbone & 38.6 & 69.9 & 5.6 & 44.0 & 69.9 & 4.2 & 30.7 & 62.2 & 11.2 & 30.3 & 64.1 & 7.2 & 37.4 & 72.9 & 10.6 & 30.4 & 74.6 & 5.9 & 46.3 & 69.6 & 11.5 \\
\multicolumn{22}{l}{\textcolor{gray}{\textit{\quad Context selectors (25\% catalog retention)}}} \\
\quad + EmbeddingGemma 300M & 34.5 & 70.5 & 5.3 & 39.3 & 68.2 & 5.8 & 25.5 & 57.1 & 9.3 & 20.9 & 59.4 & 7.7 & 44.9 & 75.1 & 7.4 & 26.0 & 73.6 & 7.2 & 45.7 & 71.1 & 6.7 \\
\quad + Gemini Embedding 2 & 41.5 & 74.1 & 3.9 & 37.2 & 67.4 & 4.3 & 29.2 & 59.2 & 7.5 & 32.2 & 66.0 & 8.6 & 45.8 & 76.1 & 7.7 & 34.2 & 77.4 & 8.1 & 40.4 & 70.3 & 7.5 \\
\quad + Jina Reranker 3.5 & 44.7 & 73.7 & 7.2 & 43.5 & 71.2 & 7.3 & 29.5 & 58.9 & 9.2 & 29.8 & 62.8 & 13.4 & 44.8 & 75.5 & 10.8 & 28.6 & 75.7 & 15.4 & 43.8 & 71.3 & 23.5 \\
\rowcolor{SysRowBlue}
+ \sys & 51.2 & 78.0 & 3.7 & 56.3 & 78.3 & 3.1 & 51.1 & 73.7 & 7.8 & 42.4 & 71.7 & 5.3 & 60.0 & 82.1 & 5.8 & 46.6 & 83.7 & 4.9 & 52.2 & 73.8 & 5.1 \\
\midrule \multicolumn{22}{l}{\textit{\sys full pipeline evaluation}} \\
\addlinespace[1pt]
gemma-4-E4B-it & 15.2 & 64.5 & 4.3 & 7.5 & 49.9 & 4.1 & 2.5 & 43.4 & 4.2 & 6.6 & 55.7 & 6.2 & 7.8 & 55.6 & 6.2 & 11.1 & 68.9 & 7.1 & 24.3 & 58.8 & 8.4 \\
\rowcolor{SysRowBlue}
+ \sys{} [encoder] & 25.2 & 67.9 & 3.6 & 9.6 & 51.9 & 4.1 & 10.3 & 55.5 & 3.7 & 12.5 & 57.0 & 3.9 & 14.8 & 57.1 & 3.7 & 13.2 & 70.3 & 3.7 & 28.9 & 64.6 & 3.9 \\
\rowcolor{SysRowBlue}
+ \sys{} [curator] & 38.2 & 79.4 & 3.4 & 11.7 & 53.9 & 3.7 & 12.9 & 60.1 & 3.5 & 13.1 & 61.2 & 4.1 & 17.9 & 58.7 & 3.8 & 14.7 & 71.7 & 5.0 & 36.2 & 70.6 & 3.9 \\
\rowcolor{SysRowBlue}
+ \sys{} [curator + post-train] & 49.3 & 86.4 & 3.1 & 25.1 & 70.0 & 3.3 & 19.2 & 64.0 & 3.6 & 22.8 & 71.7 & 4.2 & 36.2 & 72.9 & 3.8 & 21.3 & 79.6 & 3.8 & 43.3 & 76.0 & 4.0 \\
\hdashline[1pt/1pt]
\addlinespace[1pt]
+ curator w/ supervised reorganizer & 32.5 & 75.0 & 3.0 & 6.7 & 52.1 & 3.1 & 6.0 & 43.6 & 3.2 & 8.3 & 55.2 & 3.7 & 16.3 & 59.3 & 3.8 & 12.9 & 69.9 & 3.9 & 31.8 & 66.6 & 4.3 \\
\midrule \multicolumn{22}{l}{\textit{\sys context curation across other frozen LLM backbones}} \\
\addlinespace[1pt]
grok-4.1-fast-reasoning & 40.6 & 72.4 & 36.7 & 44.0 & 70.6 & 34.6 & 25.7 & 60.7 & 40.6 & 18.1 & 54.1 & 31.0 & 29.8 & 69.7 & 27.7 & 21.6 & 72.5 & 38.2 & 28.1 & 57.5 & 27.7 \\
\rowcolor{SysRowBlue}
+ \sys & 58.9 & 84.0 & 29.2 & 50.6 & 77.4 & 25.3 & 46.8 & 72.6 & 33.7 & 39.2 & 71.0 & 25.6 & 61.1 & 83.5 & 23.2 & 39.6 & 83.0 & 29.3 & 53.6 & 76.2 & 19.4 \\
\addlinespace[1pt]
grok-4.1-fast-non-reasoning & 14.8 & 58.8 & 8.0 & 21.6 & 54.7 & 8.0 & 12.6 & 52.4 & 8.0 & 18.3 & 52.4 & 9.7 & 11.8 & 45.1 & 10.0 & 3.1 & 59.6 & 10.5 & 7.7 & 29.2 & 11.9 \\
\rowcolor{SysRowBlue}
+ \sys & 46.2 & 79.7 & 1.7 & 39.0 & 72.2 & 2.1 & 22.3 & 64.7 & 2.4 & 31.9 & 70.2 & 2.5 & 29.7 & 69.9 & 3.0 & 22.2 & 77.8 & 3.5 & 26.2 & 50.3 & 3.4 \\
\addlinespace[1pt]
kimi-k2-thinking & 40.1 & 72.6 & 69.4 & 39.4 & 70.3 & 65.7 & 21.9 & 55.2 & 73.8 & 22.4 & 59.0 & 65.7 & 34.6 & 69.7 & 75.2 & 23.3 & 74.6 & 83.5 & 32.0 & 59.5 & 61.5 \\
\rowcolor{SysRowBlue}
+ \sys & 57.1 & 82.7 & 33.1 & 49.5 & 76.4 & 35.1 & 33.9 & 66.1 & 53.0 & 37.9 & 70.4 & 32.9 & 54.5 & 79.6 & 44.2 & 38.4 & 83.0 & 47.1 & 47.1 & 72.1 & 38.2 \\
\addlinespace[1pt]
glm-5.2 & 32.6 & 56.2 & 120.3 & 28.8 & 45.3 & 129.9 & 15.8 & 34.6 & 151.1 & 31.2 & 62.7 & 109.0 & 42.4 & 73.1 & 131.4 & 33.1 & 77.1 & 122.1 & 47.0 & 72.6 & 95.2 \\
\rowcolor{SysRowBlue}
+ \sys & 54.1 & 79.8 & 109.9 & 43.4 & 70.2 & 107.7 & 30.5 & 58.6 & 135.2 & 42.5 & 70.3 & 110.4 & 56.0 & 79.5 & 147.1 & 43.7 & 81.4 & 106.6 & 59.2 & 77.6 & 100.1 \\
\addlinespace[1pt]
claude-sonnet-5 & 40.8 & 74.5 & 22.5 & 36.3 & 66.4 & 27.1 & 21.6 & 55.5 & 40.0 & 31.6 & 62.0 & 31.6 & 40.5 & 70.3 & 35.6 & 30.7 & 71.7 & 47.5 & 37.7 & 69.1 & 19.6 \\
\rowcolor{SysRowBlue}
+ \sys & 54.1 & 80.7 & 21.0 & 43.5 & 72.3 & 22.4 & 34.1 & 64.2 & 33.9 & 36.0 & 65.7 & 28.4 & 52.4 & 77.6 & 31.6 & 39.8 & 82.3 & 33.5 & 54.2 & 78.4 & 20.4 \\
\addlinespace[1pt]
gemini-3.1-pro & 57.5 & 80.2 & 69.0 & 54.7 & 77.8 & 80.2 & 48.8 & 72.4 & 152.0 & 49.1 & 73.7 & 111.4 & 51.2 & 77.2 & 137.0 & 51.5 & 83.3 & 128.7 & 52.5 & 74.7 & 68.8 \\
\rowcolor{SysRowBlue}
+ \sys & 69.9 & 87.3 & 47.1 & 70.8 & 87.7 & 44.8 & 68.7 & 84.3 & 91.6 & 55.7 & 77.9 & 95.8 & 52.3 & 78.4 & 127.0 & 55.5 & 85.6 & 115.9 & 65.0 & 82.0 & 50.7 \\
\addlinespace[1pt]
claude-opus-5 & 54.0 & 79.6 & 19.1 & 43.8 & 71.8 & 20.3 & 30.0 & 57.9 & 24.3 & 31.0 & 61.7 & 20.8 & 44.4 & 73.1 & 25.4 & 34.1 & 78.4 & 24.6 & 53.7 & 77.9 & 21.7 \\
\rowcolor{SysRowBlue}
+ \sys & 54.6 & 80.9 & 19.9 & 49.2 & 76.9 & 18.7 & 41.9 & 67.6 & 21.7 & 37.4 & 66.6 & 19.9 & 55.4 & 78.1 & 24.6 & 48.6 & 83.8 & 23.6 & 64.3 & 82.3 & 19.5 \\
\bottomrule
\end{tabular}
\end{adjustbox}
\end{table*}

\subsection{Failure Analysis}
\label{app:failure_cases}

The case study in Section~\ref{subsec:main_res} follows one query that \sys
answers correctly after post-training. This appendix takes the complement: of
the queries \sys still answers incorrectly, what does it return instead, and
what separates that answer from the reference? We analyse the per-trial logs
behind the \mbox{\emph{+ \sys [curator + post-train]}} row of
Table~\ref{tab:main_results}: 700 queries (7 merchants $\times$ 100), 10 trials
each, recording the selected variant, the model's stated reason for
selecting it, and its feature reward on every trial.

\paragraph{The answers returned instead are close substitutes.}
The mean feature reward of an error trial is $0.63$, so a typical alternative
already matches most of what the query asked for. Invalid or out-of-catalog
identifiers account for $0.1\%$ of error trials. The cases below are the
recurring kinds, and each one turns on a distinction between candidates that
the catalog record describes almost identically.

\paragraph{Variants of the same product.}
$8.1\%$ of error trials across all merchants, and $27.2\%$ on PF, return the
reference's parent product under a different variant. PF sells most products in
several pack sizes, which is the setting this arises in. Consider:

\begin{quote}
\small
\refstepcounter{exquery}\label{q:pack}%
\textbf{\theexquery.} ``I'm looking for a \qreq{completely plastic-free and
PVOH-free laundry detergent} for \qcontext{my newborn twins who have severe
eczema}. I \qpref{strongly prefer the convenience of pre-measured drop-in
pods} [\ldots] However, because \qcontext{we go through so many clothes}, I
\qreq{absolutely need a package that provides at least 100 loads}. My budget
is \qreq{strictly under \$20}.''
\end{quote}

\noindent
For \ref{q:pack}, the reference is the 140-load pack of \emph{The Simplest Baby
Laundry Detergent} (\$19.99); in 8 of 10 trials \sys returns the 100-load pack
of the same product (\$16.99), with the stated reason that ``the 100 Load
variant meets the minimum load requirement and is under budget.'' Both packs
satisfy both hard constraints, and the cue that favours the larger one, going
through many clothes, is carried by a subordinate clause. The two packs are
$0.09$ FR apart, so the feature reward also treats them as close.

\paragraph{Sibling products that differ in one unscored field.}
The same structure recurs one level up, between products that are identical in
every scored field. Consider:

\begin{quote}
\small
\refstepcounter{exquery}\label{q:shade}%
\textbf{\theexquery.} ``I'm attending \qcontext{a super long outdoor wedding}
and need a \qreq{completely transfer-proof, vegan liquid lipstick} that won't
budge when I eat cake or drink champagne. I'm looking for a \qreq{striking warm
red or a deep berry shade with a matte finish}, though \qpref{I'd consider a
metallic one if it's just as smudge-proof}. My budget is \qreq{pretty strict at
under 25 dollars} [\ldots]''
\end{quote}

\noindent
For \ref{q:shade}, \sys returns the reference, \emph{Cranberry Stiletto Matte
Lip Whip}, in 6 of 10 trials, and in the other 4 returns \emph{Versailles} or
\emph{Mon Ch\'eri Matte Lip Whip}, which carry the same product type, the same
\$22.00 price and the same 23\,g weight, and are separated from the reference
only by shade name and description text (``a smudge-proof, vegan, matte liquid
lipstick in a berry-sienna hue, matching the user's request for a deep berry
shade''). The case study in Section~\ref{subsec:main_res} follows this same
discrimination on a numeric field, where post-training moves the model from
23\,g substitutes to the 15\,g reference; \ref{q:shade} is the same query
family with a colour word in place of the number.

\subsection{Catalog-Increase Protocol}
\label{app:catalog_increase}

This appendix details the evaluation protocol behind
Figure~\ref{fig:catalog_increase}, in particular how the evaluation set is
composed at each injection level.

\paragraph{Catalog construction.}
The experiment uses a single merchant, Beauty Bakerie (BB, 96 products /
115 variants; Table~\ref{tab:data}). We form the reduced initial catalog by
randomly removing half of the products \emph{within each product type}, so
that every product type survives in the reduced catalog and the type
distribution is preserved. The full \sys pipeline---encoder, curator, and
post-training---is trained once on this reduced catalog and is never retrained
thereafter. The withheld products are then reintroduced in ten cumulative
steps: the set injected at level $\ell$ is a strict superset of the set
injected at level $\ell-10$, so a product that has returned never leaves again.
At $0\%$ the model sees only the reduced catalog; at $100\%$ the entire BB
catalog has been restored, i.e.\ the catalog has doubled relative to the
catalog the model was trained on.

\paragraph{Evaluation set and reference labels.}
The same 100 evaluation queries, each run for 10 trials, are used at every
injection level; no query is added or dropped as the catalog grows. Reference
labels, however, are not constant. For the 56 queries whose original
ground-truth product falls in the withheld half, no correct answer exists in
the reduced catalog, so we re-establish a substitute reference by re-running
the self-consistency filtering and human verification stages of our data
pipeline (Section~\ref{subsec:setups}, Appendix~\ref{app:data}) against the
reduced catalog.
When such a query's original product is reintroduced, its reference label
reverts to that product. A query is counted as \emph{New-product} at a given
level once its reference label has been replaced by a reintroduced product,
and as \emph{Original-product} while its reference label is still the one it
carried at $0\%$. Group membership therefore shifts as products return, and
Table~\ref{tab:catalog_increase_composition} reports the resulting composition
at every level.

\begin{table}[h]
\centering
\scriptsize
\setlength{\tabcolsep}{4pt}
\caption{Composition of the BB evaluation set at each injection level. The
query set is fixed at 100 queries $\times$ 10 trials throughout, so the counts
below are also percentages of the evaluation set. \emph{New-product} counts
queries whose reference label has been replaced by a reintroduced product;
\emph{Original-product} counts the remainder.}
\label{tab:catalog_increase_composition}
\begin{tabular}{lrrrrrrrrrrr}
\toprule
\textbf{Products reintroduced (\%)} & \textbf{0} & \textbf{10} & \textbf{20} & \textbf{30} & \textbf{40} & \textbf{50} & \textbf{60} & \textbf{70} & \textbf{80} & \textbf{90} & \textbf{100} \\
\midrule
\# queries, Total            & 100 & 100 & 100 & 100 & 100 & 100 & 100 & 100 & 100 & 100 & 100 \\
\# queries, Original-product & 100 & 96  & 96  & 93  & 93  & 87  & 86  & 79  & 62  & 58  & 44  \\
\# queries, New-product      & 0   & 4   & 4   & 7   & 7   & 13  & 14  & 21  & 38  & 42  & 56  \\
\bottomrule
\end{tabular}
\end{table}

\paragraph{Reading the New-product panel.}
Because products are reintroduced gradually, the New-product group is very
small at low injection levels: it contains 4 queries at $10\%$ and $20\%$ and
7 queries at $30\%$ and $40\%$. The $100.0$ EM values at $10\%$ and $20\%$ are
therefore $4/4$, and the $57.1$ at $30\%$ is exactly $4/7$. These leftmost
points should be read as indicative only; the New-product trend is meaningful
from roughly $50\%$ onward, where the group reaches 13 queries or more. The
Total column is the one comparison whose denominator is constant across the
sweep, and the relative-decline figures quoted in
Section~\ref{subsec:generalizability} are computed on it.
\subsection{Feature-Reward Views of the Ablations}
\label{app:fr_views}

\begin{figure}[!h]
\centering
\includegraphics[width=6.30in]{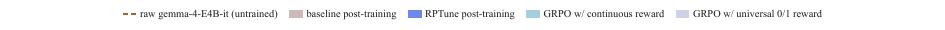}\\[-1pt]
\begin{minipage}[b]{1.52in}\centering
  \includegraphics[width=\linewidth]{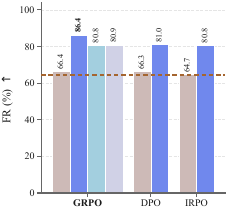}\\[-1pt]
  {\scriptsize (a) Objectives}
\end{minipage}\hfill
\begin{minipage}[b]{1.25in}\centering
  \includegraphics[width=\linewidth]{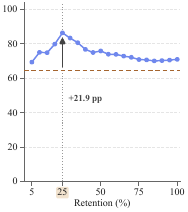}\\[-1pt]
  {\scriptsize (b) Pruning budget}
\end{minipage}\hfill
\begin{minipage}[b]{1.88in}\centering
  \includegraphics[width=\linewidth]{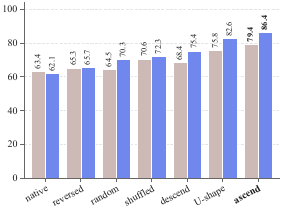}\\[-1pt]
  {\scriptsize (c) Context orderings}
\end{minipage}\hfill
\begin{minipage}[b]{1.65in}\centering
  \includegraphics[width=\linewidth]{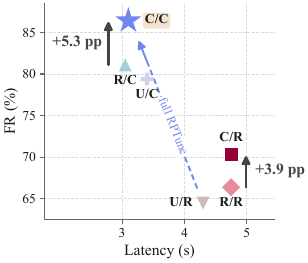}\\[-1pt]
  {\scriptsize (d) Curation quadrants}
\end{minipage}
\caption{FR view of the ablations: (a) post-training objectives
(Figure~\ref{fig:posttrain}), (b) pruning budget sweep
(Figure~\ref{fig:pruning_sweep}), (c) context layouts on Beauty Bakerie
(Figure~\ref{fig:ordering}) and (d) the curation quadrants
(Figure~\ref{fig:ablation_pareto_curation}).}
\label{fig:fr_appendix}
\end{figure}

Figure~\ref{fig:fr_appendix} gives the feature reward (FR) view of the
ablations in Section~\ref{subsec:ablation_analysis}, over the same runs and
aggregation as the main-text figures each panel mirrors; FR tracks EM
throughout. Tags in panel (d) pair the post-training catalog with the
evaluation catalog (U untrained, R raw, C curated), and its arrows give the FR
gain from curated training at a fixed evaluation context.

\subsection{Supervised Reorganizer Objective}
\label{app:supervised_reorganizer}

This appendix details the supervised control used in the reorganizer ablation of
Section~\ref{subsec:ablation_analysis}. The ablation asks whether \sys's
LLM-in-the-loop objective (Eq.~\ref{eq:reorganizer_loss}) is necessary, or whether
ordinary supervised ranking with the same capacity and the same relevance
supervision suffices.

\minihead{What is held fixed}
The supervised reorganizer differs from \sys's reorganizer only in its training
objective. It uses the same scoring function $s_i$
(Section~\ref{subsec:workflow}) on top of the same frozen encoder, updates only the
MLP parameters $\theta$, and starts from the same zero-initialized adjustment, so
training begins from the encoder ranking. It is trained on the same synthetic
queries and relevance scores $\mathcal{D}_{\text{rel}}$
(Section~\ref{subsec:training}) and parameterizes the same catalog-wide softmax
$w_\theta(c\mid q)$ of Eq.~\ref{eq:reorganizer_loss}, here with $\beta=1$. At inference time, the curated context
is built identically (top 25\%, ascending order) and passed to the same frozen
\texttt{gemma-4-E4B-it}, evaluated under the protocol of
Section~\ref{subsec:setups}.

\minihead{Objective}
We turn each query's graded relevance scores into a target distribution over the
catalog and minimize its cross-entropy with $w_\theta$:
\begin{equation}
\label{eq:supervised_loss}
p_T(c\mid q)=\frac{\exp\bigl(y(q,c)/T\bigr)}{\sum_{c'\in\mathcal{C}}\exp\bigl(y(q,c')/T\bigr)},
\qquad
\mathcal{L}_{\text{Sup}}(\theta)
=-\frac{1}{B}\sum_{q\in\mathcal{B}}\sum_{c\in\mathcal{C}}p_T(c\mid q)\log w_\theta(c\mid q).
\end{equation}
We choose this listwise form so that $\mathcal{L}_{\text{Sup}}$ and
$\mathcal{L}_{\text{Reorganizer}}$ are both cross-entropies against the same
$w_\theta$ and differ only in their target; the ablation therefore isolates the
training signal rather than the model or the loss family. The target temperature
$T$ controls how sharply $p_T$ concentrates on the most relevant products: as
$T\to 0$, all mass moves to $\arg\max_c y(q,c)$, and as $T\to\infty$, $p_T$ becomes
uniform. Since $y\in[0,1]$, $T=0.1$ turns every $0.1$ gap in relevance into an
$e$-fold ratio in target mass, so the best matches dominate while near-substitutes
retain partial credit.

\minihead{Supervised versus LLM-in-the-loop gradients}
The two objectives update the scores in different ways. For a query $q$,
\[
\frac{\partial \mathcal{L}_{\text{Sup}}}{\partial s_i}
\propto \frac{1}{\beta}\Bigl(w_\theta(c_i\mid q)-p_T(c_i\mid q)\Bigr),
\qquad
\frac{\partial \mathcal{L}_{\text{Reorganizer}}}{\partial s_i}
\propto -\frac{1}{\beta G}\sum_{g\in\mathcal{V}_q} A_g\Bigl(\mathbb{1}[a_g=c_i]-w_\theta(c_i\mid q)\Bigr).
\]
The supervised gradient is dense and fixed: every product moves toward its label
mass at every step, independently of what the downstream LLM does with the
resulting context. In the LLM-in-the-loop gradient, when all $G$ rollouts are
valid, the standardized advantages sum to zero, the $w_\theta$ terms cancel, and
only products that the frozen LLM actually selected from the current context
receive gradient: selections that beat their group average are promoted and the
others are demoted. The RL signal thus concentrates on the LLM's observed behavior
on the contexts the reorganizer currently builds, for example demoting a plausible
but less relevant product that the LLM tends to select, whereas the supervised
target does not depend on the LLM at all.

\minihead{Training details}
We minimize $\mathcal{L}_{\text{Sup}}$ with AdamW (learning rate $10^{-3}$, linear
warmup over the first 10\% of steps followed by linear decay) on minibatches of
$B=128$ queries, as in RL training (Appendix~\ref{app:curves}), for 20 epochs over
each merchant's training queries. We use $T=0.1$ and the final (epoch-20)
checkpoint for every merchant; both were fixed in advance rather than tuned on the
evaluation queries. Supervised training requires no LLM calls and is therefore
considerably cheaper than RL training.

\minihead{Results}
The dashed row of the \texttt{gemma-4-E4B-it} block in
Table~\ref{tab:full_results} reports the per-merchant results. Averaged over the
seven merchants, the supervised reorganizer reaches 16.4\% EM and 60.2\% FR,
essentially matching the encoder-only curator (16.4\% EM, 60.6\% FR) and falling
4.3 EM and 4.9 FR points short of \sys's LLM-in-the-loop reorganizer (20.7\% EM,
65.1\% FR). It improves over the encoder on only three merchants (BB, WB, and PF)
and does not improve on the other four (OON, HK, WCP, and FF), whereas \sys's
reorganizer improves over the encoder on all seven merchants and achieves higher EM
than the supervised reorganizer on each of them. Relevance supervision alone
therefore does not yield consistently better contexts; the gains of \sys's
reorganizer come from optimizing against the downstream LLM's selections.

\end{document}